\documentclass[twocolumn,showpacs,preprintnumbers,amsmath,amssymb,nofootinbib,prd]{revtex4}

\usepackage{graphicx}
\usepackage{dcolumn}
\usepackage{bm}
\usepackage{epsf}
\usepackage[applemac]{inputenc}
\usepackage{amsmath}
\usepackage{amssymb}

\begin{document}

\title{Probing non-Gaussianities in the CMB on an incomplete sky using surrogates}

\author{G. Rossmanith$^1$, H. Modest$^1$, C. R\"ath$^1$, A. J. Banday$^{2,3}$, K. M. G\'{o}rski$^{4,5}$ and G. Morfill$^1$}
\affiliation{	$^1$ Max-Planck-Institut f\"ur extraterrestrische Physik, Giessenbachstr. 1, 85748 Garching, Germany \\
		$^2$ CNRS; IRAP; 9 Av. colonel Roche, BP 44346, F-31028 Toulouse cedex 4, France \\
		$^3$ Université de Toulouse; UPS-OMP; IRAP;  Toulouse, France \\
		$^4$ Jet Propulsion Laboratory, California Institute of Technology, Pasadena, CA 91109, USA \\
                $^5$ Warsaw University Observatory, Aleje Ujazdowskie 4, 00 - 478 Warszawa, Poland}
                
\date{\today}

\begin{abstract}
We demonstrate the feasibility to generate surrogates by Fourier-based methods for an incomplete data set. This is performed for the case of a CMB analysis, where astrophysical foreground emission, mainly present in the Galactic plane, is a major challenge. The shuffling of the Fourier phases for generating surrogates is now enabled by transforming the spherical harmonics into a new set of basis functions that are orthonormal on the cut sky. The results show that non-Gaussianities and hemispherical asymmetries in the CMB as identified in several former investigations, can still be detected even when the complete Galactic plane ($|b| < 30^\circ$) is removed. We conclude that the Galactic plane cannot be the dominant source for these anomalies. The results point towards a violation of statistical isotropy.
\end{abstract}


\maketitle

\label{firstpage}


\section{Introduction}

The search for primordial non-Gaussianities in the Cosmic Microwave Background (CMB) is one of the most important yet challenging tasks in modern cosmology. Any convincing detection of intrinsic non-Gaussianities as well as their characteristics and scaling behaviour would directly support or reject different models of inflation, and therefore affect a fundamental part of the standard cosmological model.

The currently still favoured inflationary model is single-field slow-roll inflation \cite{Guth81,Linde82,Albrecht82}, which should result in (nearly) Gaussian and isotropic temperature fluctuations of the CMB. However, deviations from Gaussianity, preferred directions, and other kinds of asymmetries have been repeatedly detected \cite{Eriksen04b, Eriksen07,Hansen07,Dickinson09,Hoftuft09,Hansen09,Rossmanith09,Raeth09,Raeth10,Vielva10,Pietrobon10,Paci10,Cayon10,Copi10}, already questioning the simplest picture of inflation. It is under discussion, if these asymmetries are connected to foreground influences \cite{Copi09, Naselsky11}, which appear particularly in the direction of the Galactic plane.


For investigations of CMB data sets, e.g. WMAP data, the analysis of Fourier phases has proven to be a useful method \cite{Chiang03,Coles04,Naselsky05,Chiang07}, since all potential higher order correlations, which directly point to non-Gaussianities, are contained in the phases and the correlations among them. The method of surrogate maps with shuffled Fourier phases \cite{Raeth09,Raeth10} represents one way of analysing the phases. Originally, this idea stems from the field of time series analysis \cite{Theiler92, Schreiber98, GVK341772518, Gliozzi10} and describes the construction of data sets, so-called surrogates, which are similar to the original, except for a few modified characteristics. The validation of these characteristics in the original data can then be tested by comparing them to the set of surrogates with appropriate measures. The method used in \cite{Raeth09,Raeth10} tests the hypothesis that the coefficients $a_{lm} = \left| a_{\ell m} \right| e^{i \phi_{lm}}$ of the Fourier transform of the temperature values $T(\theta,\phi)$ have independent and uniform distributed phases $\phi_{\ell m} \in [-\pi,\pi]$ calculated for the complete sphere $S$. The phases of the original map are shuffled, which can be done within some previously chosen interval of interest, $\Delta \ell = [\ell_1,\ell_2]$, or simply for the complete range $\Delta \ell = [2,\ell_{max}]$ for some given $\ell_{max}$. Every realisation of this shuffling results in a new set of $a_{\ell m}$´s, which then represents (after transforming back) one surrogate map. Note that every surrogate still has by construction exactly the same power spectrum as the original map. If the original map contained any phase correlations, these are now destroyed due to the shuffling. Thus, any detected differences between the original and a set of surrogate maps reveals higher order correlations and therefore deviations from Gaussianity.


One major problem in CMB analyses is the treatment of the Galactic plane, which strongly influences the microwave signal. It is possible to cut out the foreground affected regions \cite{Gold10}, but this procedure itself can affect the subsequent analyses. When applying a sky cut, orthonormality of the spherical harmonics no longer holds on this new incomplete sky, which leads to a coupling of the $a_{\ell m}$`s, making a naive phase shuffling impossible. However, one can transform the spherical harmonics into a new set of harmonics, which forms an orthonormal basis on the incomplete sky \cite{Gorski94a,Gorski94b,Mortlock02}, where phase manipulation can then take place again.

The problem of incomplete data also occurs in time series analysis by means of surrogates. Here, gaps can be overcome e.g. by the use of simulated annealing for reproducing the autocorrelation function \cite{Schreiber99, Schreiber00}. Still, the quality of surrogates constructed with this method seems to be questionable, since it is not ensured that no phase correlations are induced. 

In this work, we combine the cut sky methods with phase shuffling, thus enabling investigations by means of surrogates on an incomplete sky. Our method can also be extended for use on incomplete data sets in general.


\begin{figure}
\centering
\includegraphics[width=4.27cm, keepaspectratio=true, ]{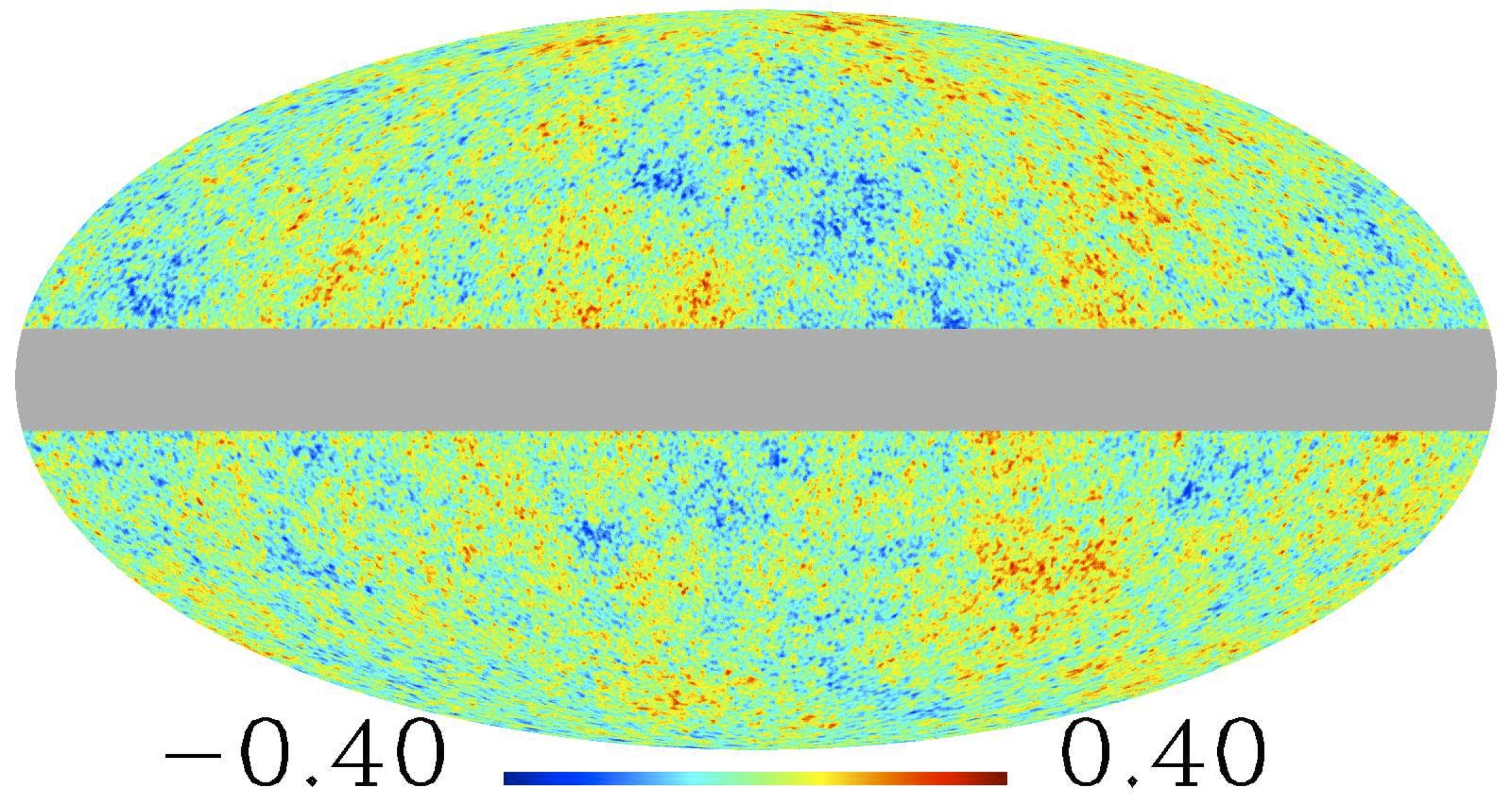}
\includegraphics[width=4.27cm, keepaspectratio=true, ]{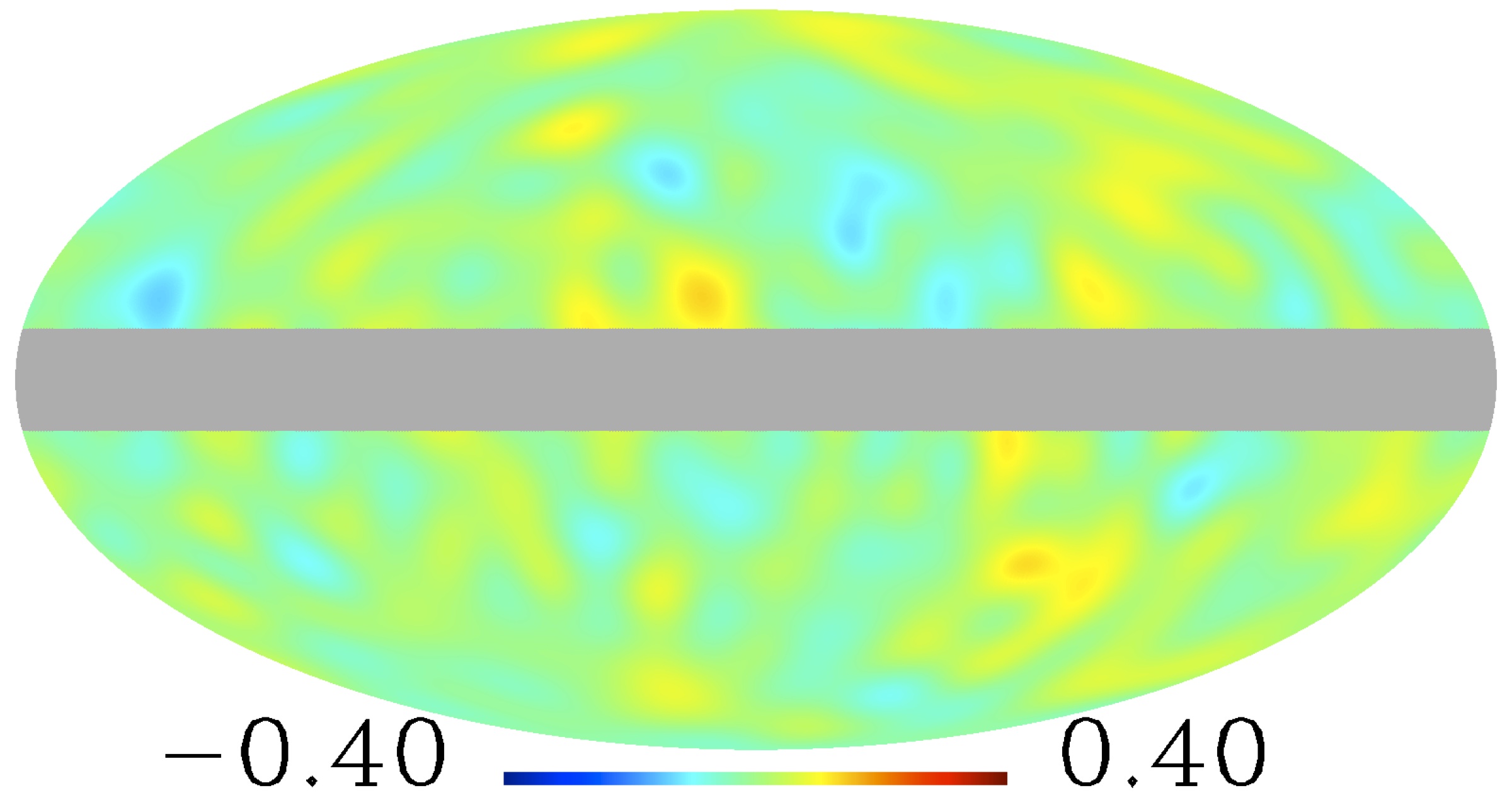}
\includegraphics[width=4.27cm, keepaspectratio=true, ]{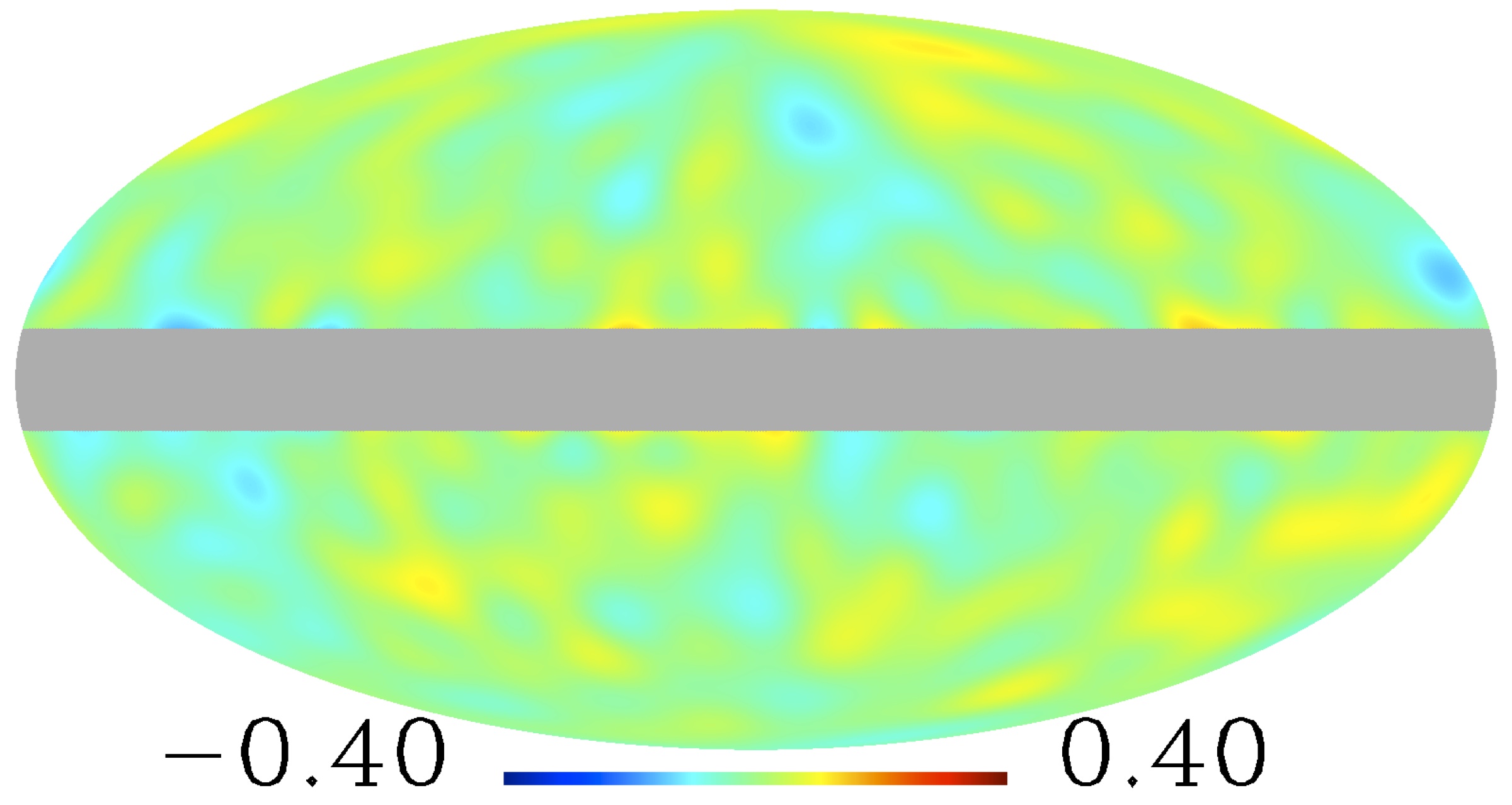}
\includegraphics[width=4.27cm, keepaspectratio=true, ]{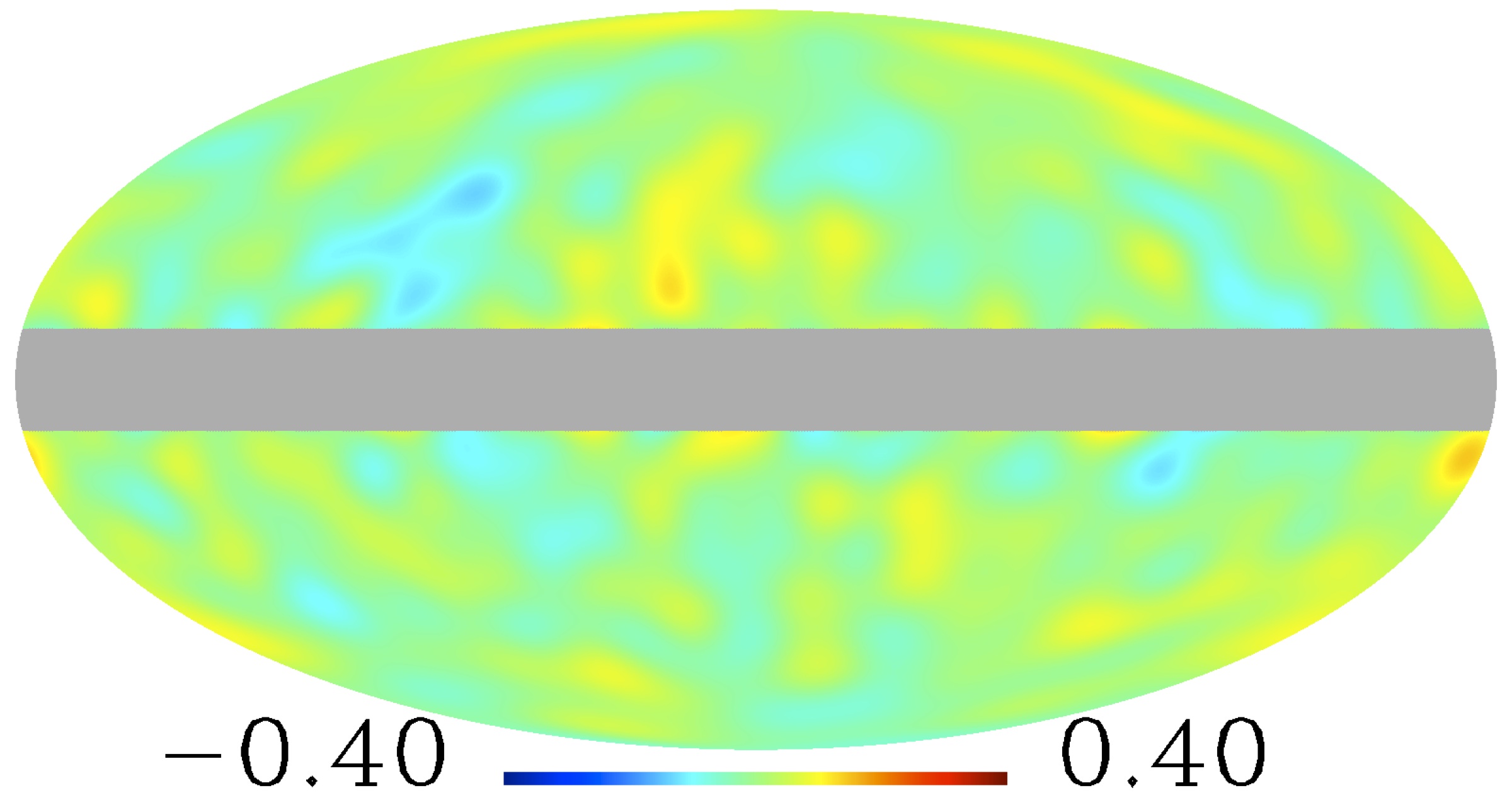}
\caption{Example for the method of surrogates on a cut sky: A simulated CMB map in its original form (upper left), after reconstruction with cut sky harmonics corresponding to the sky cut $|b| < 10^\circ$ and the limitation $\ell_{max}=20$ (upper right), and after two realisations of a phase shuffling of these harmonics for the complete range of $\Delta \ell = [2,20]$ (lower row).} \label{Beispielsurrogate}
\end{figure}

\section{Methods}

\subsection{Cut Sky Surrogates}

On a complete sphere $S$, an orthonormal basis is given by the spherical harmonics $Y_{\ell m}(s)$ with $\ell \ge 0 , -\ell \le m \le \ell$ and $s \in S$. Let the number of harmonics be limited by some given $\ell_{max} \in \mathbb{N}^+$. Now, for any map $f(s) = \sum\nolimits_{\ell,m}^{\ell_{max}} a_{\ell m} Y_{\ell m}(s) ,\ \forall \ s \in S$ with $a_{\ell m} \in \mathbb{C}$, and for any new incomplete sky $S^{cut}$, we want to know the corresponding $a^{cut}_{\ell m}$ and $Y^{cut}_{\ell m}$ for representing the map on the remaining regions of the sphere: $f(s) = \sum\nolimits_{\ell,m}^{\ell_{max}} a^{cut}_{\ell m} Y^{cut}_{\ell m}(s) ,\ \forall \ s \in S^{cut}$, with $Y^{cut}_{\ell m}$ being orthonormal on $S^{cut}$ and thus $a^{cut}_{\ell m}$ being unique. For real valued spherical harmonics, this was performed in \cite{Gorski94a} and \cite{Gorski94b}, and later on extended in \cite{Mortlock02}. The methods presented there can be easily adapted to complex valued spherical harmonics as well.

At first, we define the vectors
\begin{alignat*}{2}
& Y(s) && := [ Y_{0,0}(s) , Y_{1,0}(s) , Y_{1,1}(s) , ... , Y_{\ell_{max}, \ell_{max}}(s) ]^T, \\
& a && := [ a_{0,0} , a_{1,0} , a_{1,1}  , ... , a_{\ell_{max}, \ell_{max}} ]^T
\end{alignat*}
containing all harmonics and coefficients with $m \geq 0$, respectively, of the given map on the complete sphere. Both have a length of $i_{max}:=(\ell_{max}+1)(\ell_{max}+2)/2$. Analogously, we define $Y^{cut}(s)$ and $a^{cut}$. Our objective is to determine two transformation matrices $B_1, B_2 \in \mathbb{C}^{i_{max} \times i_{max}}$, that fulfil the following equations:
\begin{alignat}{2}
& Y^{cut}(s) &&= B_1 \ Y(s) \label{Ziel1} \\
& a^{cut} &&= B_2 \ a \label{Ziel2}
\end{alignat}
To identify them, we need to define the coupling matrix
\[ C:= \int_{R}Y(s)Y^*(s)d\Omega \]
as well as analogously its counterpart $C^{cut}$, with $R$ being a given region on the sphere. Hereby, $Y^*$ denotes the hermitian transposed of $Y$. When working with a pixelised sky, one uses a sum over the pixels of $R$ instead of the integral. For $R=S^{cut}$, an orthonormal set of harmonics $Y^{cut}_{\ell m}$ needs to fulfill the condition $C^{cut}=I_{i_{max}}$, with $I_{i_{max}}$ being the unit matrix of size $i_{max}$. We can use equation (\ref{Ziel1}) on $C^{cut}$ to change this condition to $B_1CB_1^* = I_{i_{max}}$. It is possible to apply different matrix decompositions to obtain $C = A A^*$ with $A \in \mathbb{C}^{i_{max} \times i_{max}}$. Consequently, the above equation now reads as $(B_1A)(B_1A)^* = I_{i_{max}}$ and offers the simple solution $B_1 = A^{-1}$.

\begin{figure}
\centering
\includegraphics[width=8.5cm, keepaspectratio=true, ]{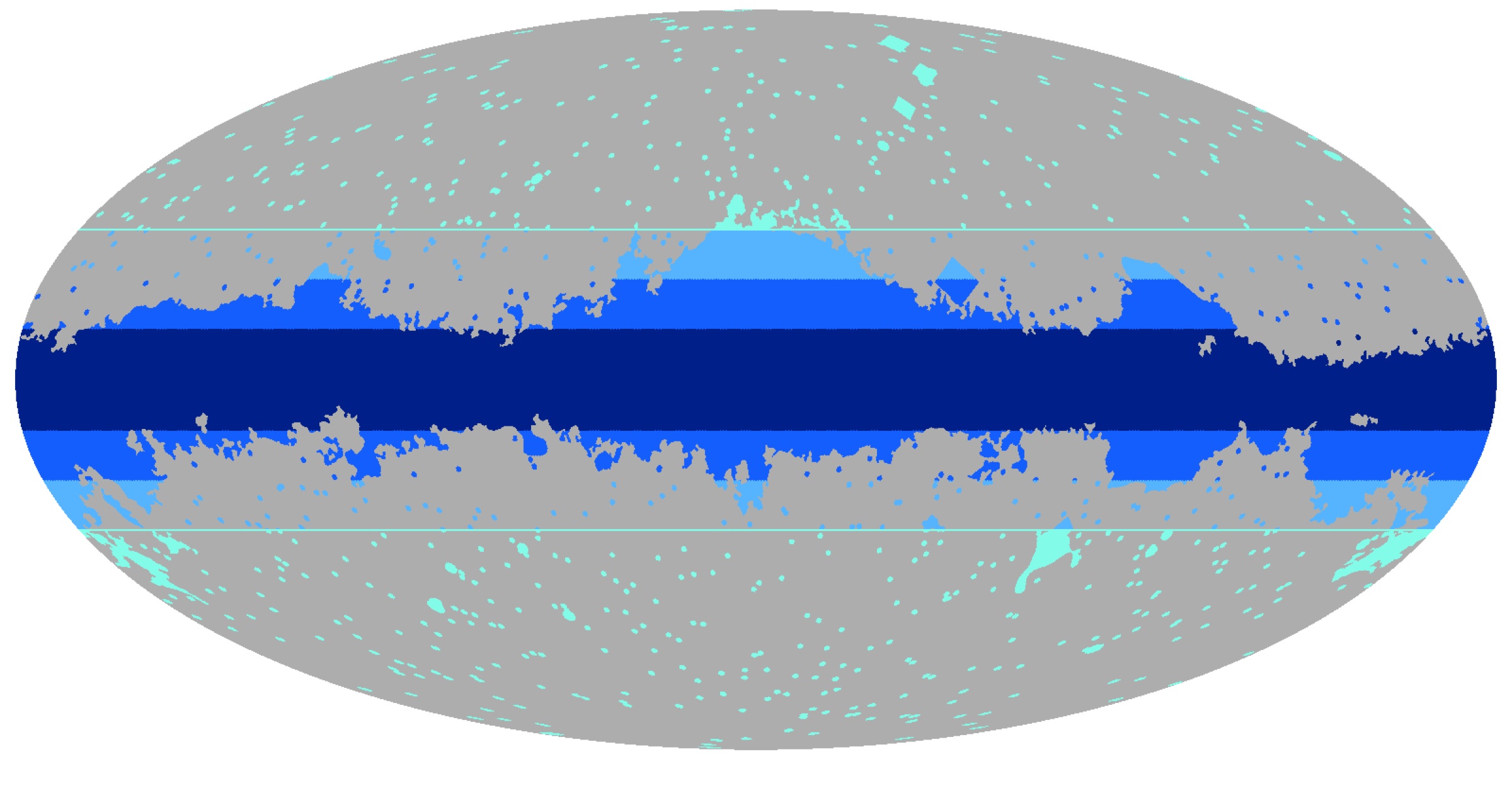}
\caption{The latest KQ75-mask provided by the WMAP-team \cite{Gold10}, that cuts out the highly foreground affected regions, compared to the three central latitude cuts applied in this work. The smallest cut is $\pm 10^\circ$ (dark blue), which removes already a large amount of the highly foreground affected regions of the Galactic plane but presumes nearly all non-affected regions. The second, $\pm 20^\circ$ (blue), removes more regions, while the largest, $\pm 30^\circ$ (light blue) cuts out almost the entire Galactic Plane. Only minor point sources (turquoise) remain.} \label{MaskenSpezialPlot}
\end{figure}

For the evaluation of $B_2$, let us recall that the coefficient vector $a$ can be expressed by
\[ a = \int_S \overline{Y}(s)f(s)d\Omega\]
or, respectively,
\[a^{cut} = \int_{S^{cut}} \overline{Y}^{cut}(s)f(s)d\Omega \ . \]
Inserting (\ref{Ziel1}) and the expression $f(s)=a^TY(s)$ into the latter leads to $a^{cut} = \overline{B}_1C^Ta$. Now we use the above matrix decomposition again and obtain $a^{cut} = \overline{B}_1(AA^*)^Ta = A^Ta$. Thus, it follows that $B_2 = A^T$.

To obtain the $a^{cut}_{\ell m}$ and $Y^{cut}_{\ell m}$ with $m < 0$, we make use of the following equations, that hold for full sky and that we assume to be valid also on incomplete skies:
\[ Y^{cut}_{\ell , -m} = (-1)^{\left| m \right|} \overline{Y}^{cut}_{\ell m}\]
and 
\[ a^{cut}_{\ell , -m} = (-1)^{\left| m \right|} \overline{a}^{cut}_{\ell m} \ . \]
For the sky cuts and $\ell$-ranges used throughout our study, the cut sky harmonics were tested and confirmed to be orthogonal.

For constructing $C=AA^*$ as above, one can make use of different matrix decomposition methods. Since the coupling matrix $C$ is hermitian and can be treated as positive definite for low $\ell$ by construction, a Cholesky decomposition is applicable. This is the easiest and fastest way, although numerical problems only allow usage for lower $\ell_{max}$ \cite{Mortlock02}. Another possibility is the eigendecomposition (ED): We obtain $C = VWV^*$, with the columns of V containing the eigenvectors, and W being diagonal and containing the eigenvalues of $C$. Because of the properties of the coupling matrix, these values are real and positive, allowing therefore a simple decomposition of $W$ by taking the square root of every element, $W=W^{1/2}(W^*)^{1/2}$. Thus, we obtain $A=VW^{1/2}$. Since $C$ is hermitian, the ED is formally similar to a singular value decomposition (SVD), which is also applied in this paper, with the eigenvalues corresponding to the singular values. For both the ED and the SVD we apply a householder transformation similar to \cite{Mortlock02} to make $A$ lower triangular. For the Cholesky decomposition, this is already the case by definition. Thus, due to equation (\ref{Ziel2}), it is ensured that the mono- and dipole contributions of the underlying maps -- often considered as non-cosmological -- are kept separate from the $\ell \geq 2$ modes.

With the help of the new cut sky harmonics $Y^{cut}_{\ell m}$, we can now generate the surrogates on a cut sky $S^{cut}$ as well. Similar to above, we shuffle the phases $\phi^{cut}_{\ell m}$ of the cut sky coefficients $a^{cut}_{\ell m}$, which is in this work performed for the full cut sky range $\Delta \ell^{cut} = [2,\ell_{max}]$. We obtain new sets of $a^{cut}_{\ell m}$´s, which are transformed back to pixel space to form the cut sky surrogate (CSS) maps. Figure \ref{Beispielsurrogate} shows as an example a simulated CMB map, its reconstruction with cut sky harmonics and two corresponding cut sky surrogates. As we did in the case of a complete sphere, we now search for deviations between the original data as well as its surrogates. However, one has to take care about the above mentioned properties. While the uniform distribution still holds for $\phi_{\ell m}^{cut}$, the single phases in the sets are no longer independent from each other due to equation (\ref{Ziel2}). In other words, the cut sky transformation induces phase correlations to the underlying map. To account for these systematic effects, we create for each of the input maps $N_{FSS}=20$ full sky surrogate (FSS) maps as explained above, with $\ell_{max} = 1024$ and by shuffling the phases within $\Delta \ell = [2,1024]$. By comparing the results of the surrogate analysis for an input map and its FSS, we evade systematically induced phase correlations, since the original map as well as the FSS are subject to the same cut sky transformation process. Thus, we only search for additional signatures possibly contained in the phases.

Even though the input map might feature induced phase correlations, it has to be pointed out that the cut sky transformation is reasonable. Only with a new set of cut sky harmonics, analysis techniques that depend on the orthonormality (like the method of surrogates) are feasible on an incomplete sky. The transformation into the new regime only induces a phase coupling to the data set, whose consequences can be eliminated as explained above. The method of cut sky surrogates itself is firmly constructed.

We construct the cut sky harmonics for three different central latitude sky cuts, that remove $|b| < 10^\circ$, $20^\circ$ and $30^\circ$ of latitude in the centre of the maps. While the smallest cut ($|b| < 10^\circ$) removes already a large amount of the highly foreground affected regions but retains nearly all non-affected regions, the largest ($|b| < 30^\circ$) excludes almost the entire Galactic plane, with only minor point sources remaining. This is illustrated in figure \ref{MaskenSpezialPlot}. We choose an upper bound of $\ell_{max} = 20$ and set $a_{\ell m} = 0$ for $\ell > \ell_{max}$. To check for consistency with \cite{Raeth09,Raeth10}, we applied the cut sky formalism also to the complete sphere with no points excluded. To compare the different matrix decomposition methods, all three approaches (Cholesky, ED, SVD) were applied. For every sky cut, the phases of the coefficients $a^{cut}_{\ell m}$ were shuffled a hundred times to generate $N_{CSS} = 100$ cut sky surrogates for each input map. The same was done for the corresponding FSS maps. 


\begin{figure}
\centering
\includegraphics[width=8.5cm, keepaspectratio=true]{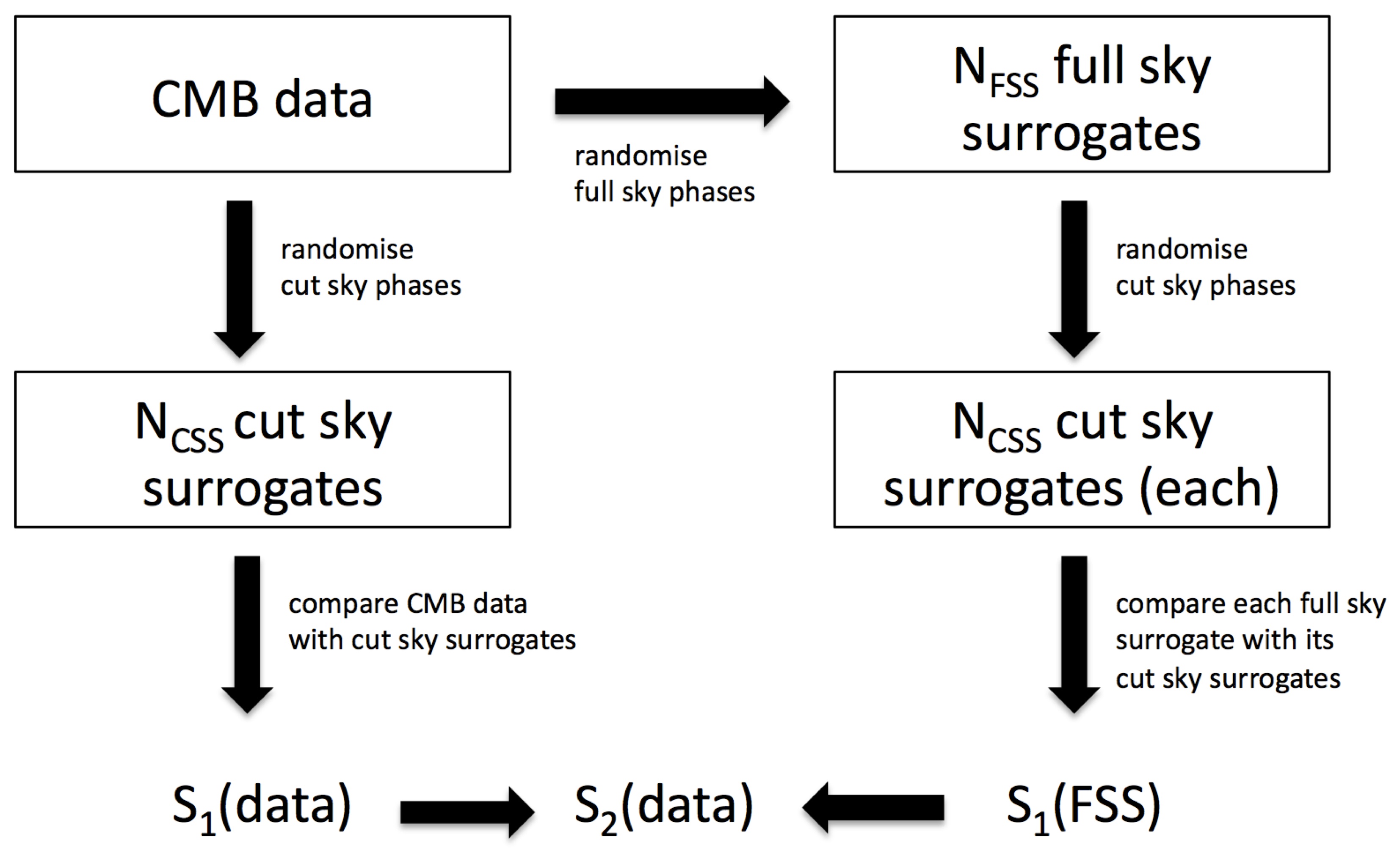}
\caption{Sketch of the method of surrogates on a cut sky.} \label{SketchalsBild}
\end{figure}

\subsection{Measures for comparison}

In general, the comparison of the original data and its surrogate maps can be accomplished with any higher order statistics. For our study, we chose the \textit{scaling index method} (SIM) \cite{Rossmanith09, Raeth10} as well as \textit{Minkowski functionals} \cite{Eriksen04b,Michielsen01} as test statistics. 

The SIM is a local measure that is able to detect structural characteristics of a given data set by estimating its local scaling properties. Briefly, the temperature anisotropies $T(\theta,\phi)$ are transformed to variations in radial direction around the sphere, therefore leading to a point distribution $\vec{p}_i$, $i = 1,\dotsc, N_{pix}$, in three-dimensional space. Then, the weighted cumulative point distribution $\rho(\vec{p}_i,r)$ is calculated for every point $\vec{p}_i$ and a freely chosen scaling parameter $r$. Since we will only investigate the large scales in our study, we choose the free parameter $r$ to be $r_{10} = 0.25$ (in the notation of \cite{Rossmanith09}), which is appropriate for these scales. Eventually, the scaling indices are obtained by calculating the logarithmic derivative of $\rho(\vec{p}_i,r)$ with respect to $r$.

The three Minkowski functionals measure the behaviour of a given map with respect to different threshold values $\nu$. The fraction of the sky where the temperature value is larger than $\nu$ is denoted as the excursion set $R(\nu)$, its smooth boundary is identified by $\partial R(\nu)$, and $da$ and $dl$ describe the surface element of $R(\nu)$ and the line element of $\partial R(\nu)$, respectively. Then, we can define the three Minkowski functionals as
\begin{align*}
&M_{area}(\nu) =\int_{R(\nu)}da \\
&M_{perim}(\nu) =\int_{\partial R(\nu)}dl \\
&M_{euler}(\nu) =\int_{\partial R(\nu)}dl \kappa \ , \\ 
\end{align*}
with $\kappa$ being the geodesic curvature of $\partial R(\nu)$. For more details, we refer to \cite{Eriksen04b,Michielsen01}. Eventually, we sum up over all thresholds with the help of the appropriate cut sky surrogates by means of a $\chi^2$-measure
\[ \chi_\bullet^2 = \sum_{\nu} \left[ \left( M_\bullet^{map}(\nu) - \langle M_\bullet^{CSS}(\nu) \rangle \right) / \sigma_{M_\bullet^{CSS}(\nu)} \right]^2\]
for $M_{area}(\nu)$, $M_{perim}(\nu)$ and $M_{euler}(\nu)$, respectively.

For computational reasons, the resolution of the input maps in the corresponding HEALPix scheme \cite{Gorski05, Healpix} was chosen to be $N_{side} = 256$ for the scaling index analysis and $N_{side} = 64$ for the Minkowski functionals. By testing several subsets, we assured ourselves that the results are only marginally affected when choosing a lower resolution.


\subsection{Statistical interpretation}

The results for the different maps of both the scaling indices and the Minkowski functionals are then evaluated in terms of \textit{rotated hemispheres}: For 768 different angles we rotate the underlying maps and calculate the $\sigma$-normalised deviations
\[ S_1(Y) = \left( Y^{map} - \langle Y^{CSS} \rangle \right) / \sigma_{Y^{CSS}}\]
of the pixels included in the new upper hemisphere between the input map and its cut sky surrogates, by means of the measure $Y$. In our case, $Y=\langle \alpha \rangle, \sigma_{\alpha}, \chi^2_{area}, \chi^2_{perim}, \chi^2_{euler}$, with $\langle \alpha \rangle$ and $\sigma_\alpha$ being the mean and the standard deviation of the scaling index response $\alpha(\vec{p}_i,r)$, respectively. The result is then shown as colour-coded pixel, whose centre is pierced by the z-axis of the respective rotated reference frame (see \cite{Raeth09, Rossmanith09, Raeth10}). To separate traces of possibly intrinsic phase correlations from those induced by the transition to incomplete sky, and thus to account for systematic effects, we calculate the statistics
\[ S_2(Y) = \left( S_1^{data}(Y) - \langle S_1^{FSS}(Y) \rangle \right) / \sigma_{S_1^{FSS}(Y)}\]
for comparing the results of $S_1(Y)$ for the original and the full sky surrogate maps. Figure \ref{SketchalsBild} illustrates the concept of this statistical interpretation of the method of cut sky surrogates.

\begin{figure}
\centering
\includegraphics[width=4.27cm, keepaspectratio=true, ]{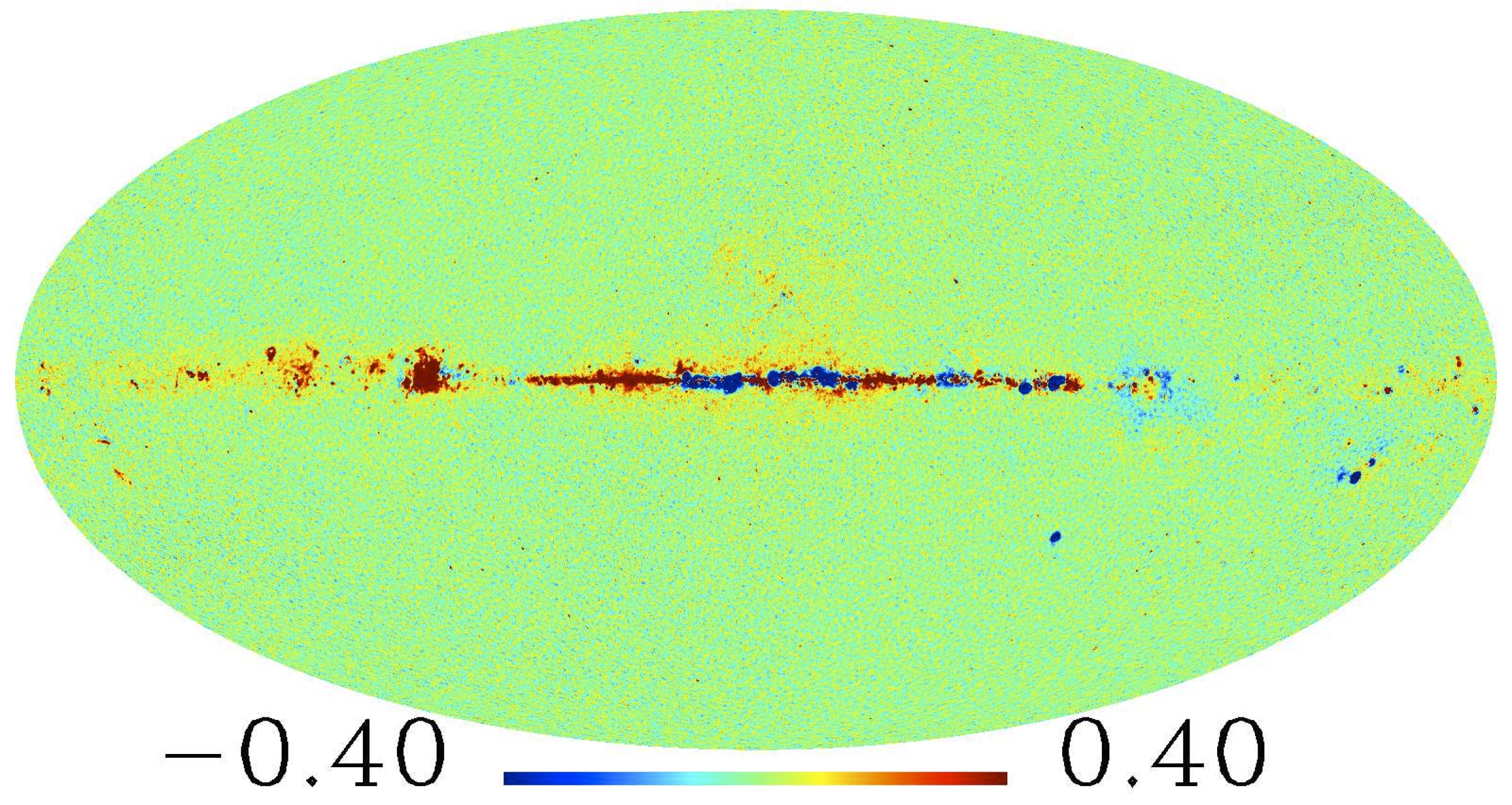}
\includegraphics[width=4.27cm, keepaspectratio=true, ]{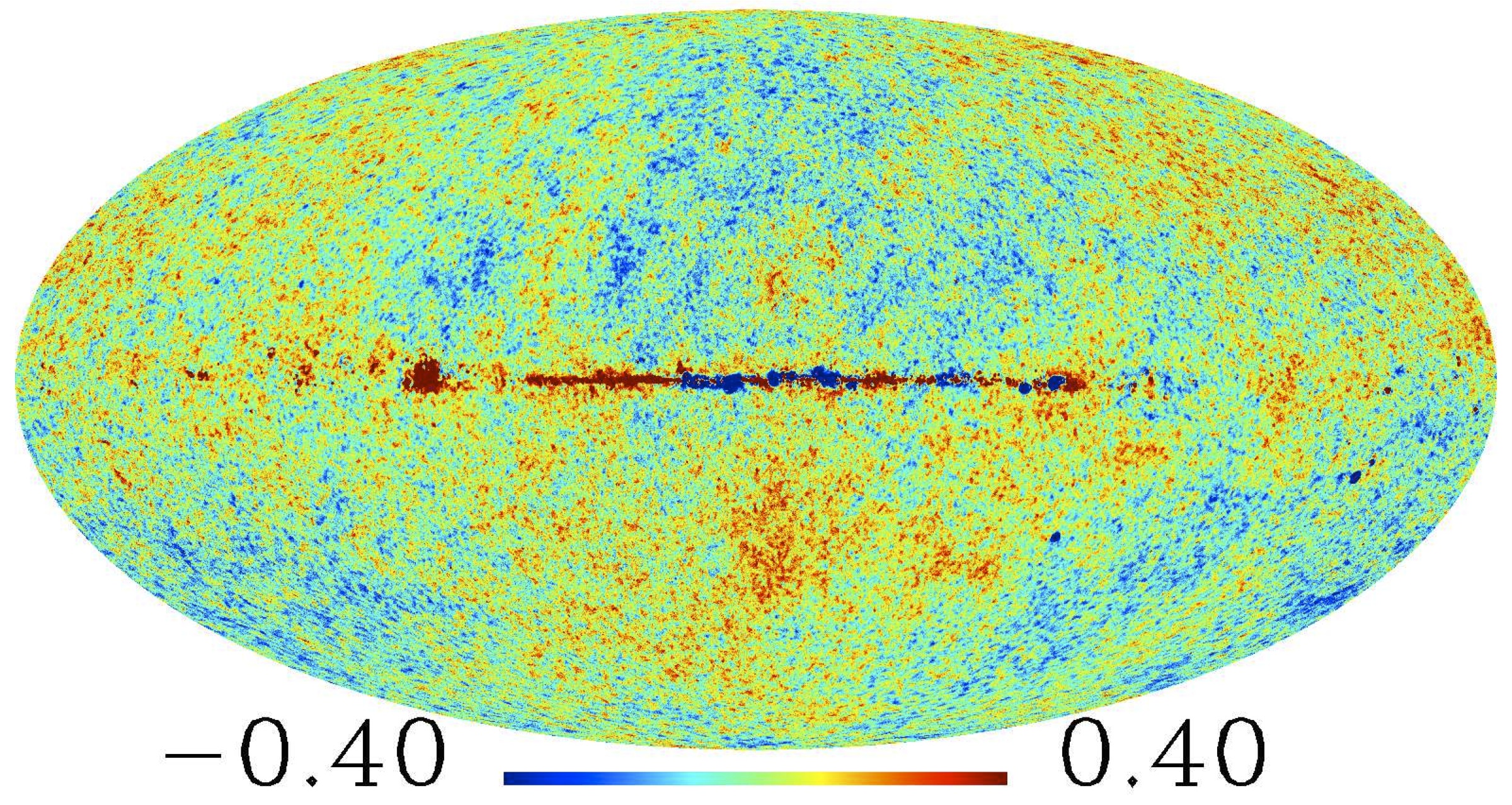}
\caption{The foreground residuals resulting from the subraction of the WMAP ILC map from the full seven-year foreground reduced coadded VW-band in its original form (left) and added to a Gaussian simulation (right). The latter represents the map that is used in section \ref{validationsectionSub1}.} \label{TonyTestMapResiduals}
\end{figure}

An important advantage of the rotated hemispheres $S_2(Y)$ is the possibility to identify not only a global difference between the original map and its surrogates, but also the directions where these differences occur. Thus, one can make a statement about possible deviations from statistical isotropy. One example for such anisotropies are asymmetries, that have been detected in numerous previous analyses as mentioned above. A difference between a $S_2(Y)$ map that detected asymmetries and a statistical isotropic map is reflected in a difference in the empirical distributions: For the latter, one would expect a histogram with a narrow range of values for $S_2(Y)$ that peaks around zero. But for the first, one expects a broader spectrum and possibly more than one peak, since on the map there are at least two areas with different values. 
One measure that exactly accounts for the broadness of histograms is the discrete entropy $H(S_2(Y))$. We define it as
\[ H(x) = - \sum_{i=1}^{N_{bins}} p_i \ln p_i \ , \]
with $p_i$ being the empirical probability (that is the relative frequency) of the realisations of $x$ to occur in the bin $i$, with $N_{bins}$ describing the number of bins in a chosen underlying interval.

The entropy $H(S_2(Y))$ reaches its maximal value for a uniform distribution, which would correspond to a $S_2(Y)$ map that consists of all different values of the chosen range, which again would imply that there is no statistical isotropy at all. The minimum of $H(S_2(Y))$ is zero. This occurs for the special case of a completely isotropic map with a distribution where only for one bin $i$ the empirical probability $p_i$ differs from zero. Thus, a statistical anisotropic map, e.g. one with a strong asymmetry, results in a higher value for $H(S_2(Y))$ than an isotropic one.


\begin{figure}
\centering
\includegraphics[width=4.27cm, keepaspectratio=true, ]{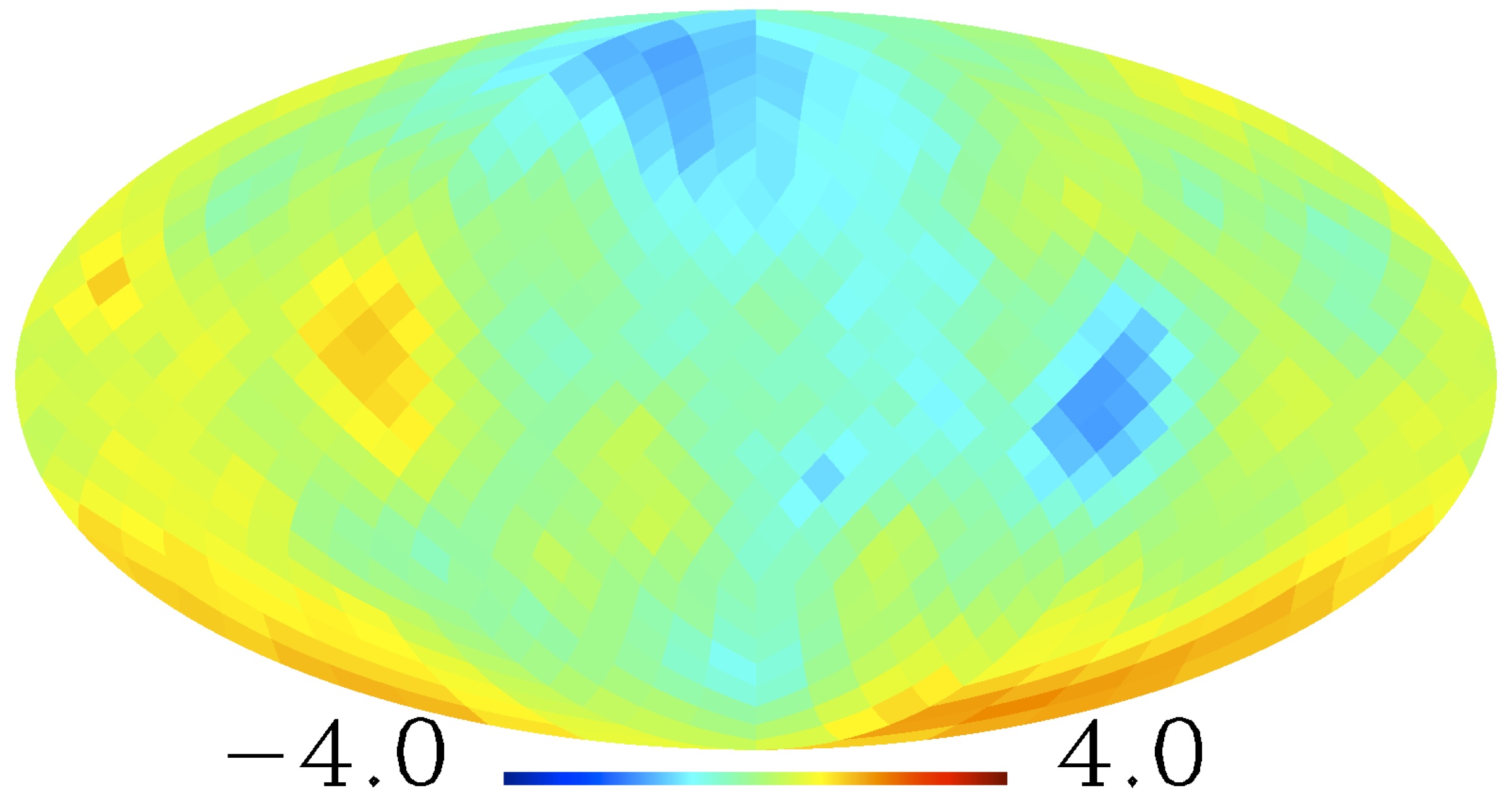}
\includegraphics[width=4.27cm, keepaspectratio=true, ]{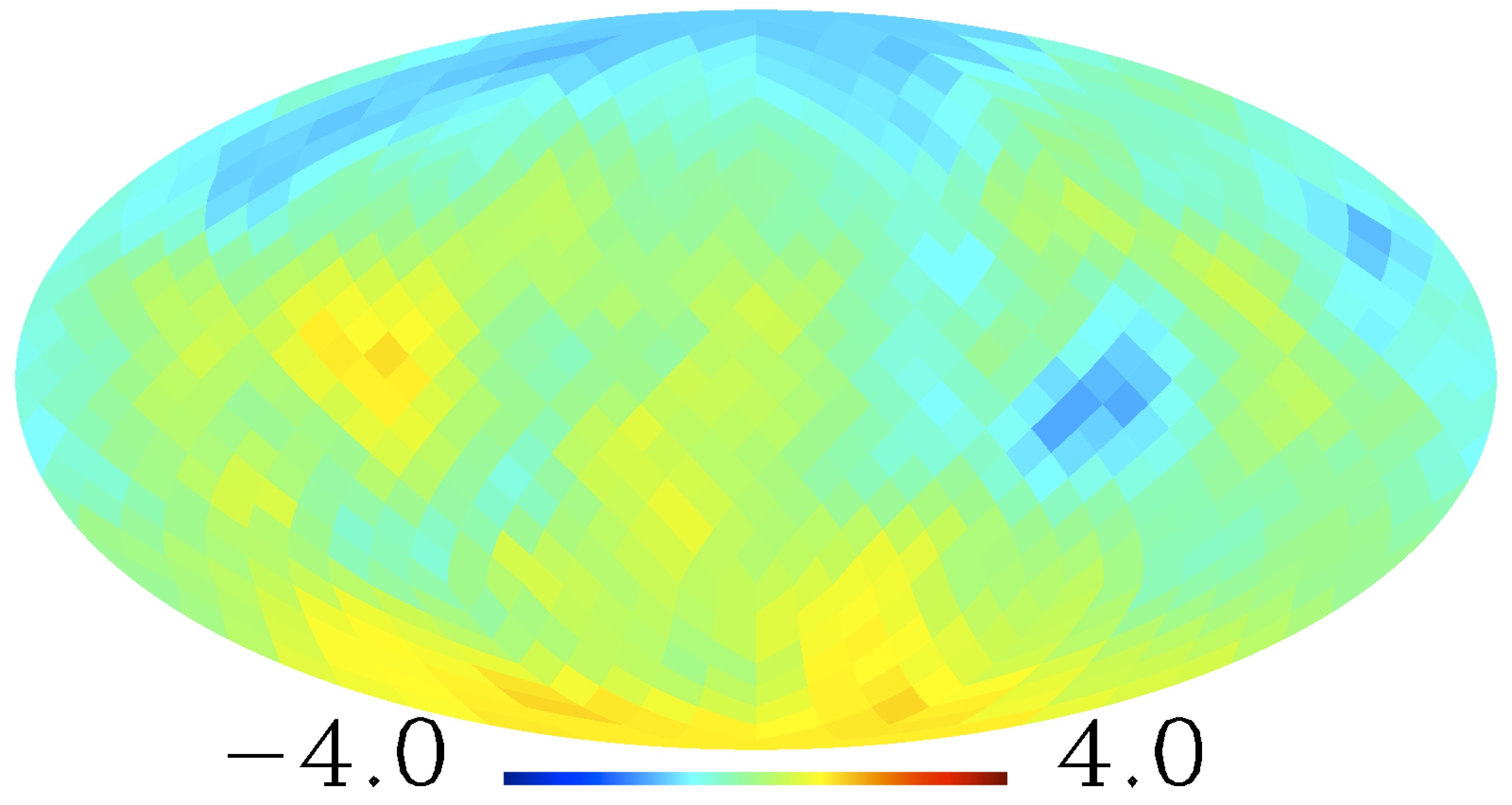}
\includegraphics[width=4.27cm, keepaspectratio=true, ]{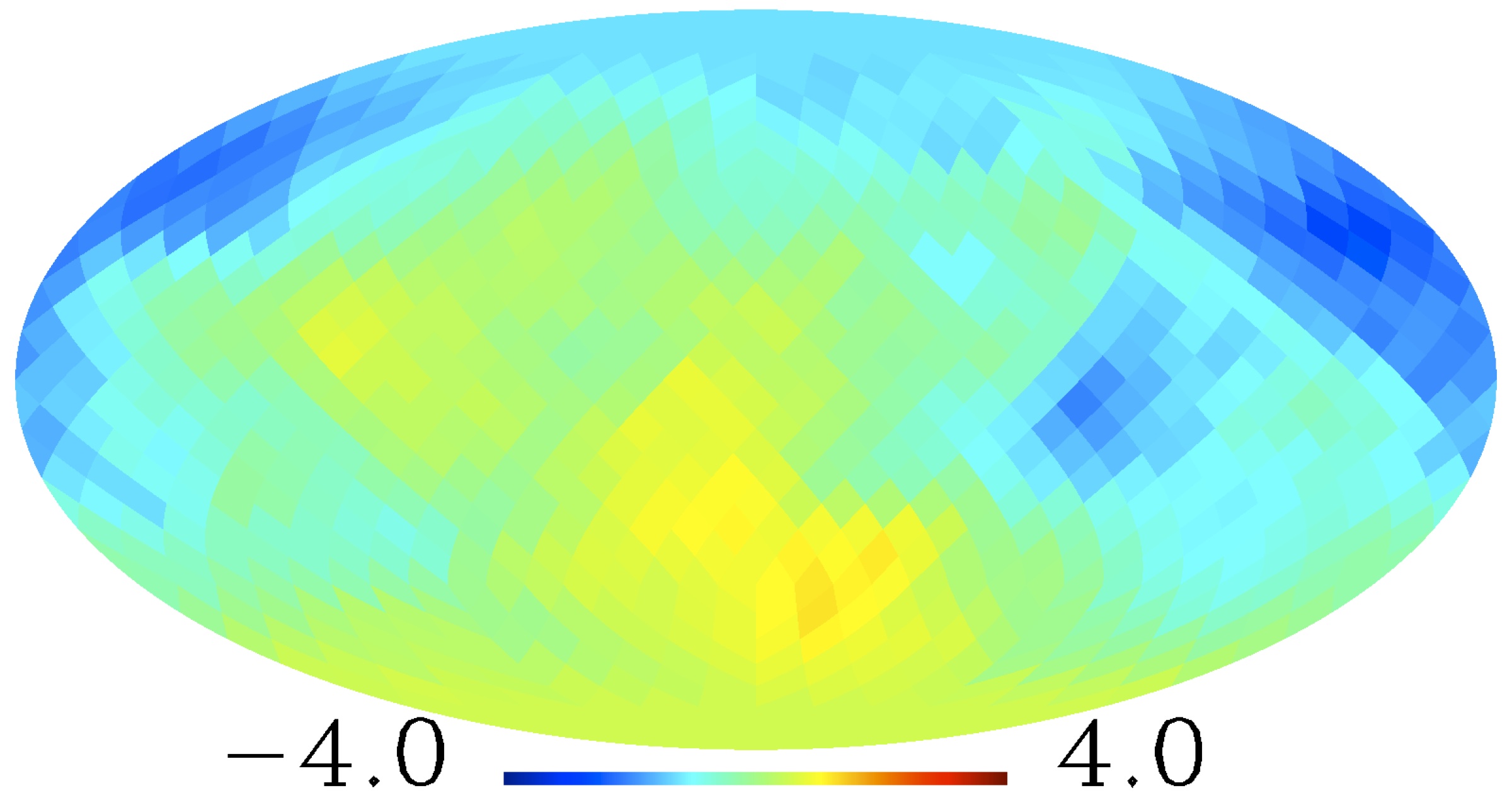}
\includegraphics[width=4.27cm, keepaspectratio=true, ]{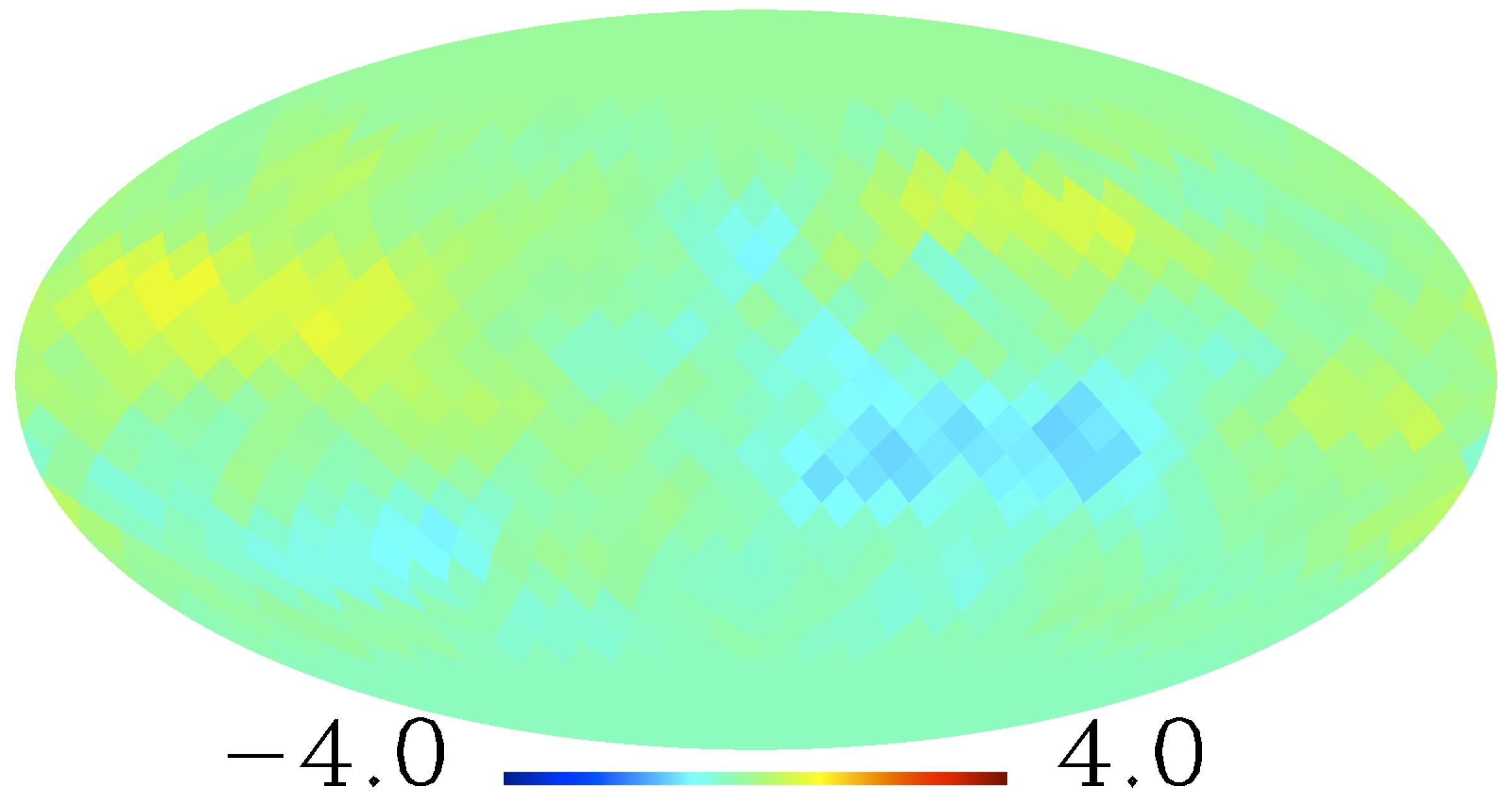}
\caption{The $\sigma$-normalised deviations $S_2(\sigma_\alpha)$ comparing a simulated map and its 20 full-sky surrogates for the complete sphere (upper left) and the three different central latitude sky cuts $|b| < 10^\circ$ (upper right), $|b| < 20^\circ$ (lower left), and $|b| < 30^\circ$ (lower right), that were constructed by means of the singular value decomposition.} \label{Heikeplots}
\end{figure}

\section{Validation}\label{validationsection}

\subsection{Underlying Data Sets}\label{validationsectionSub1}

To test the new approach, we generate two types of Gaussian simulations of the coadded VW-band of the WMAP satellite via a noise-weighted sum. The difference between these two is simply that for the one type we add Gaussian noise with a particular variance, given by $N_{obs}$, for every pixel of the sphere and for the other we do not. The procedure is the same as in \cite{Rossmanith09}, but note that we now apply the more recent WMAP 7-year parameters. We include the simulation without the noise, since only there one can be absolutely sure that the results are not distorted by any influences coming from the input map: The map is created by only using a given power spectrum and completely random phases, and should therefore be resistant against phase shuffling in a statistical sense. In addition to these two clean simulations, we applied the cut sky surrogate approach to another simulated Gaussian map, to which we added typical foreground residuals that are still present after the template cleaning of the WMAP data. With the help of this second simulation, one can examine the impact of possible aliasing effects due to the chosen $\ell_{max}$, that could be caused by strong foregrounds which are cut out. The required residuals were computed by subtracting the WMAP ILC map from the full seven-year foreground reduced coadded VW-band. However, when performing this subtraction, one has to bear two things in mind: First, not only the coadded VW-band but also the ILC map contains influences from the Galactic plane, although with a lower amplitude. Second, the resulting map is slightly noisier compared to the original foreground cleaned WMAP band-wise maps. Still, both remarks should not have a significant impact on the intention of the test. The residuals resulting from the subraction as well as the corresponding simulation with the typical foreground pattern are shown in figure \ref{TonyTestMapResiduals}.

\subsection{Results}\label{ValidationResults}

\begin{figure}
\centering
\includegraphics[width=4.27cm, keepaspectratio=true, ]{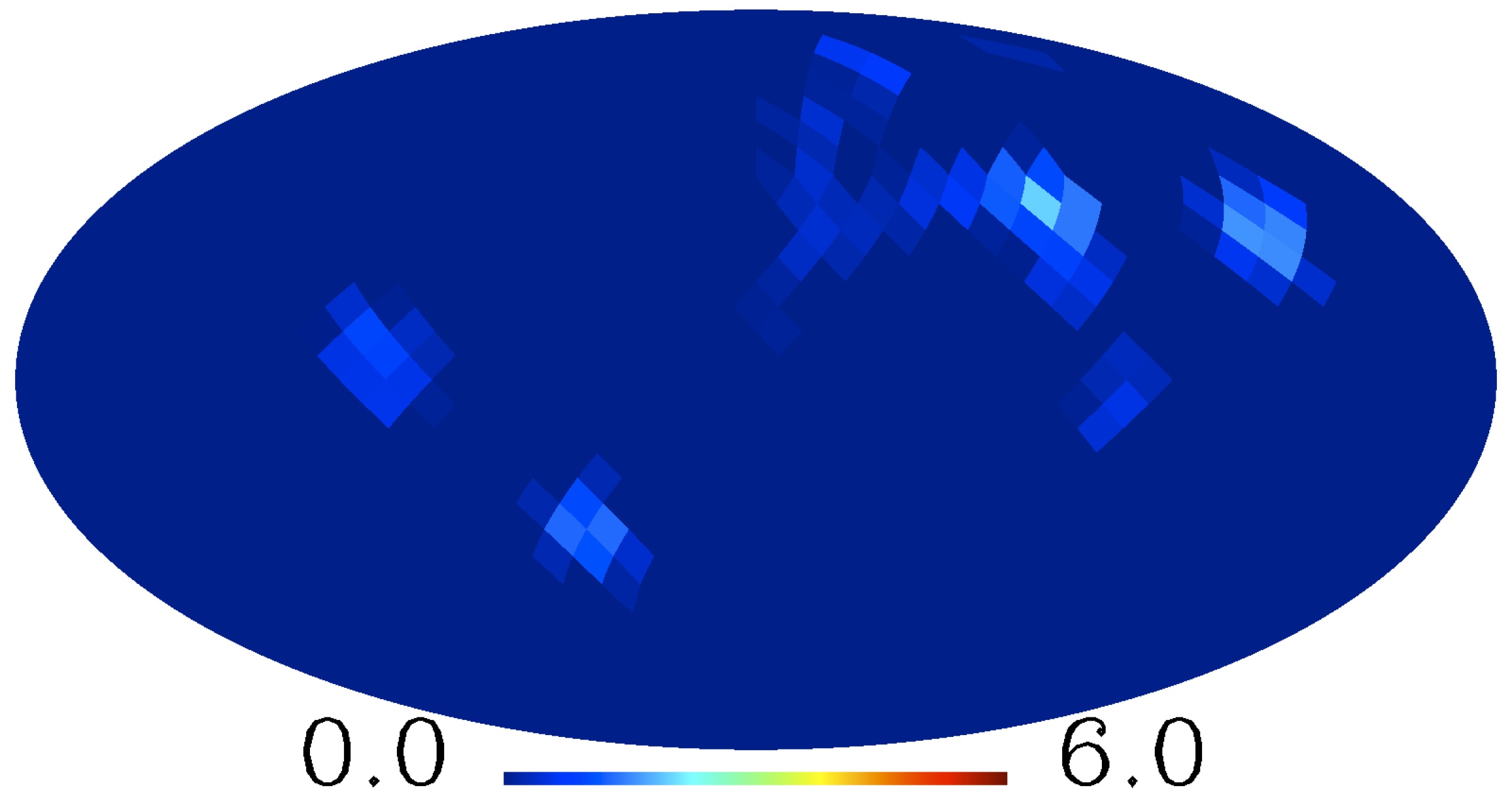}
\includegraphics[width=4.27cm, keepaspectratio=true, ]{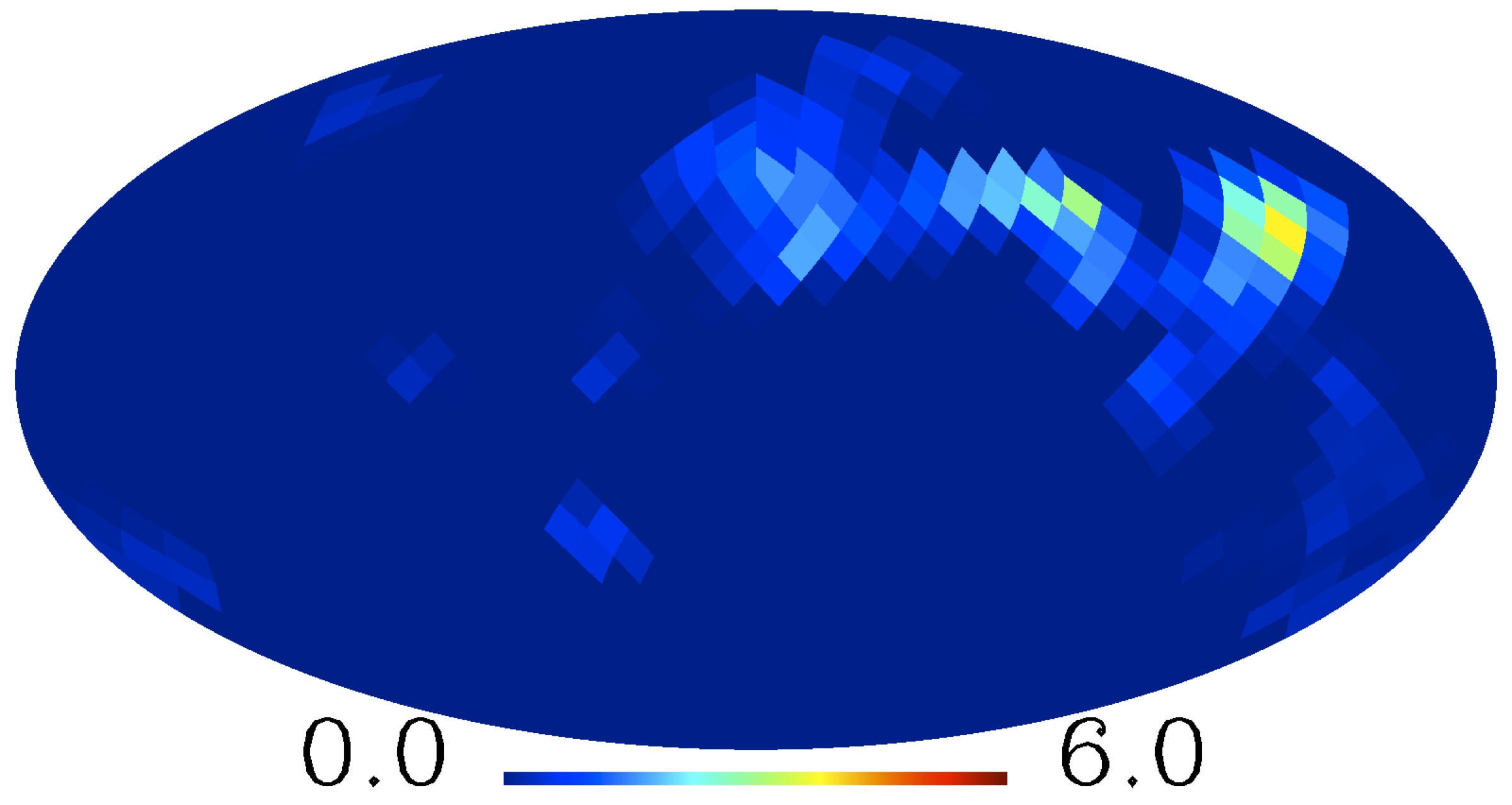}
\includegraphics[width=4.27cm, keepaspectratio=true, ]{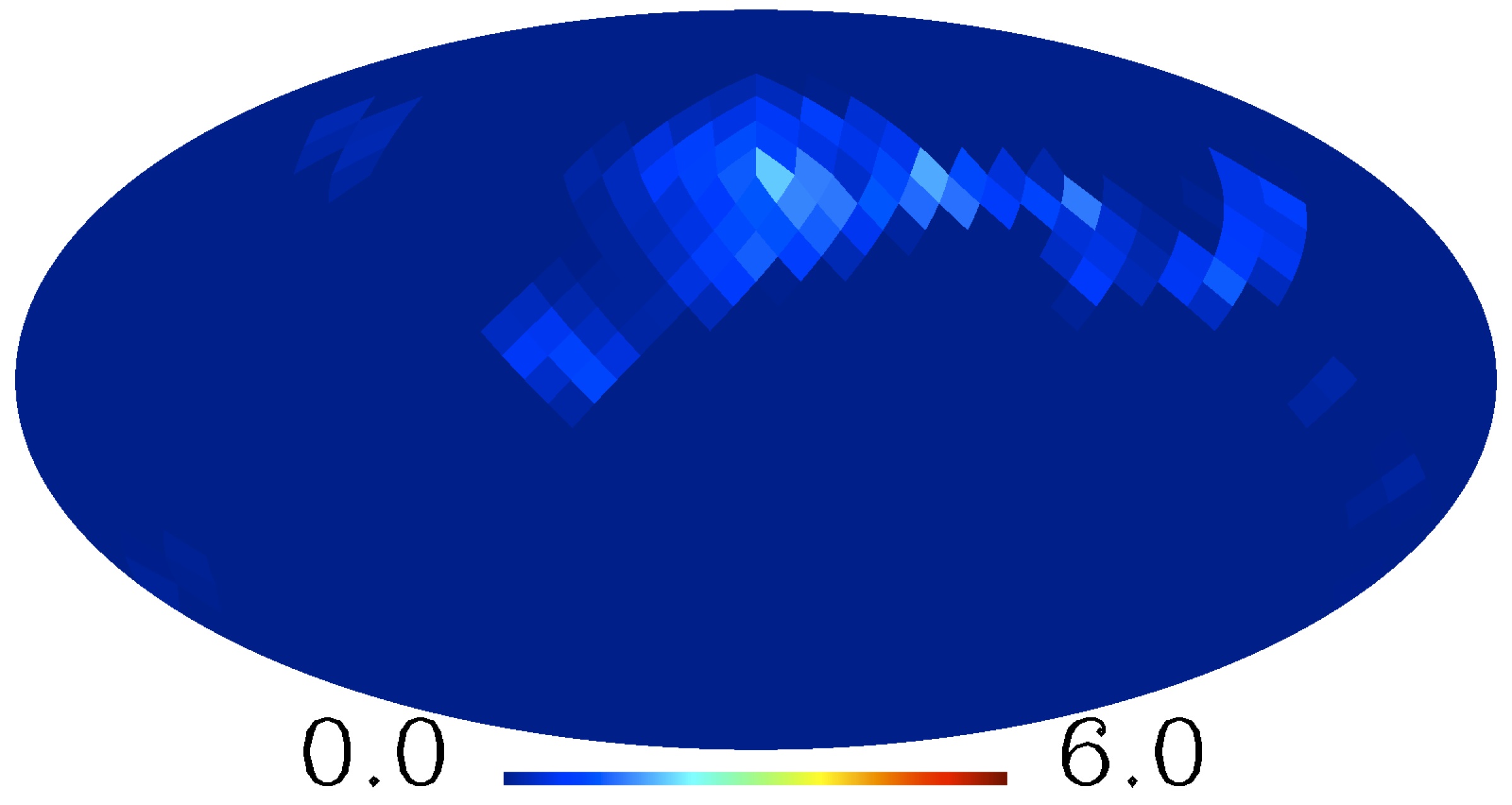}
\includegraphics[width=4.27cm, keepaspectratio=true, ]{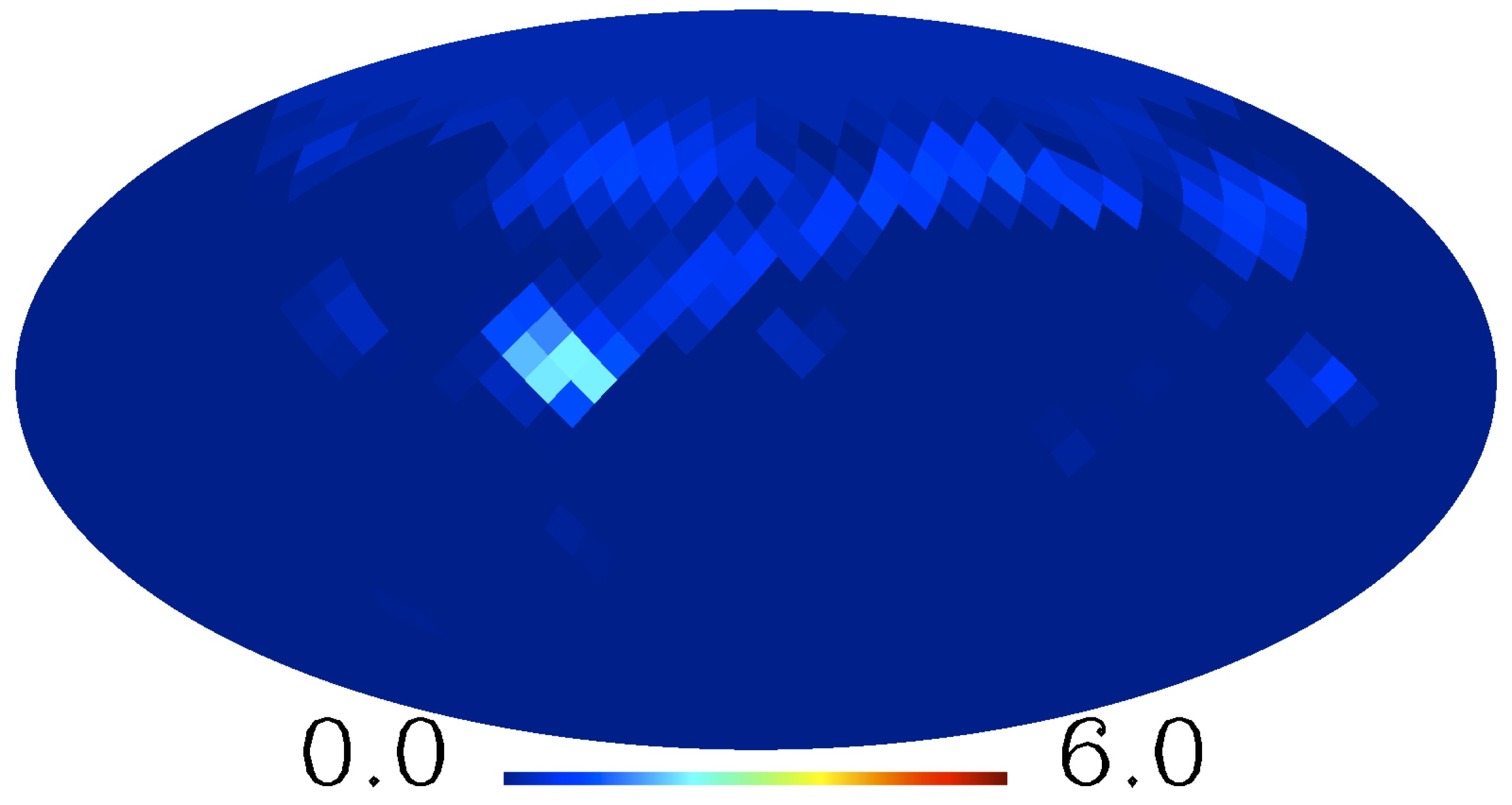}
\caption{Same as figure \ref{Heikeplots} but for $S_2(\chi_{area}^2)$. Note the different colour scaling.} \label{Heikeplots2}
\end{figure}

For the clean simulated Gaussian maps, the significances $S_2(\sigma_\alpha)$ and $S_2(\chi_{area}^2)$ calculated for the rotated hemispheres are illustrated in the figures \ref{Heikeplots} and \ref{Heikeplots2}. The differences between the original and the FSS maps are insignificant (nearly always $S_2(Y) <  2.5$) for the complete sphere and all sky cuts. This holds for the results of the scaling indices ($Y=\langle \alpha \rangle, \sigma_{\alpha}$) as well as for the Minkowski functionals ($Y=\chi^2_{area}, \chi^2_{perim}, \chi^2_{euler}$). There are only marginal differences when one adds the Gaussian noise to the map, which shows that these noise fluctuations alone are not responsible for non-Gaussianities or anisotropies on the large scales. The results of the assembled map are very similar as well, except for the full sky, where phase correlations are obviously present. These results clearly demonstrate the validity of the approach. Although phase correlations were induced by the cut sky transformation, these are subtracted away by using $S_2(Y)$. The results of the assembled map show that the impact of aliasing effects, even when strong foregrounds are present in the Galactic plane, is negligible.

Only minimal differences were detected for the three used matrix decomposition methods, that are likely to be due to the random shuffle of the phases. When going to larger $\ell$-ranges or more irregular sky cuts, this technical part of the investigation will become more important, especially for making the transition to the cut sky possible.

\begin{figure}
\centering
\includegraphics[width=8.5cm, keepaspectratio=true, ]{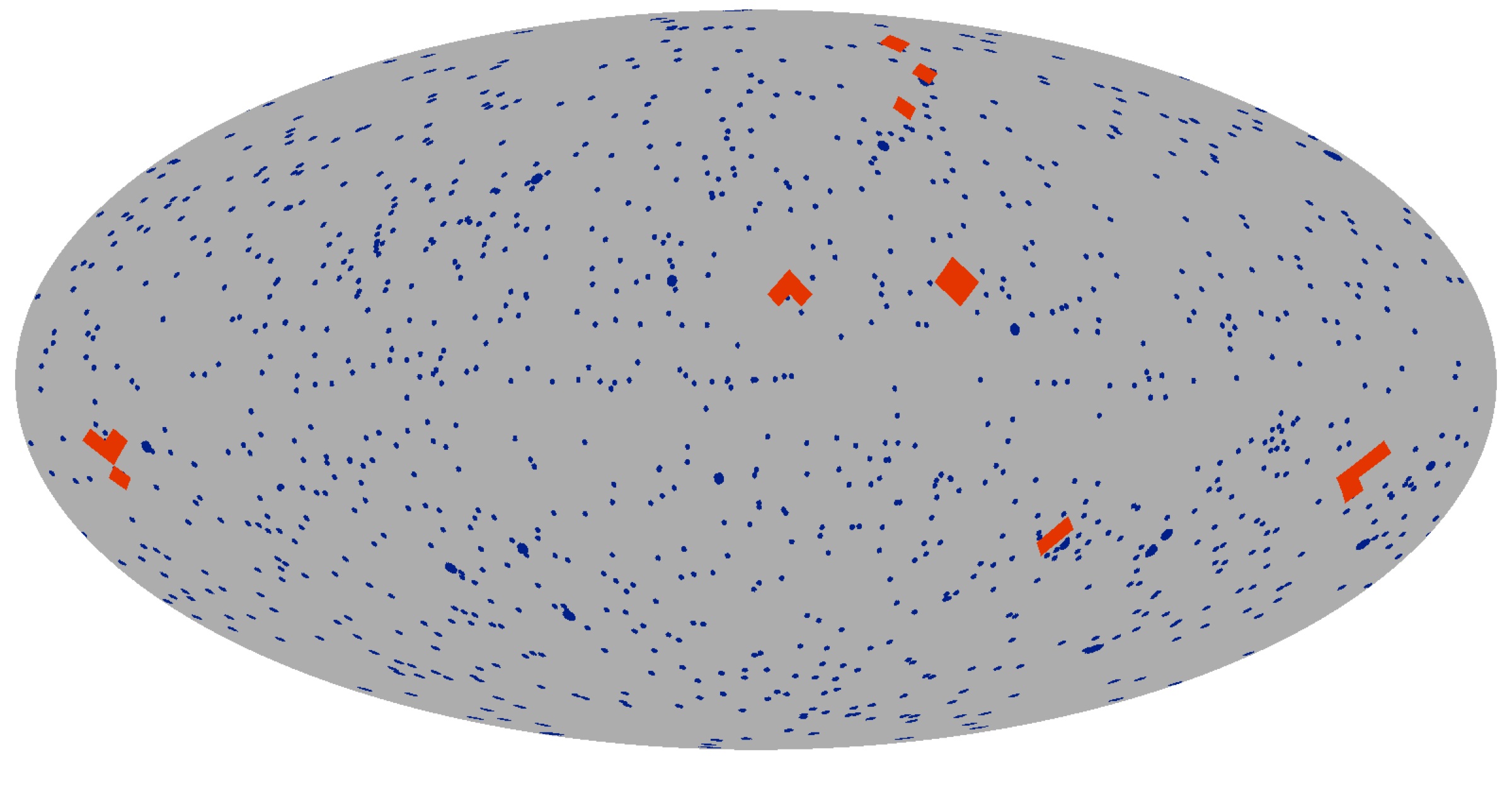}
\caption{The temperature point source mask of the WMAP team in its original (blue spots) and in its extended form (additional red regions), used to obtain the VW7p and the VW7pe data map, respectively.} \label{DieTonyMasken}
\end{figure}

\section{Application to WMAP data}

\subsection{Underlying Data Sets}

For the application of the cut sky method to observations, we make use of several data maps. Two of them are linear combinations of the different frequency bands and based on the WMAP results \cite{Jarosik10, Lambda}: First, the 7-year Internal Linear Combination map (ILC7) provided by the WMAP team \cite{Gold10} and second the 5-year needlet based ILC map (NILC5) \cite{Delabrouille09}. In addition to these two data sets, we examine three maps based on the 7-year co-added VW-band: For the first (VW7m), the galactic plane and several point sources of the co-added VW-band map are cut out by means of the KQ75-mask \cite{Jarosik10}. These empty regions are then filled with Gaussian noise, whose mean and standard deviation correspond to the values of the remaining map, while preserving the spatial noise patterns. As above, this procedure is similar to \cite{Rossmanith09}. The second map (VW7p) is obtained by applying the same method but cutting out the point sources outside the Galactic Plane only. The temperature point source mask of the WMAP team is used for this purpose \cite{Jarosik10}. Since some of the point sources correspond to larger sources that might not always be completely masked out, it can be advantageous to extend the previous mask with the help of the high-latitude regions of the WMAP processing mask \cite{Jarosik10}. This leads to the third VW-band based map (VW7pe) used in this paper. The regions that were cut out to obtain the VW7p and VW7pe maps are illustrated in figure \ref{DieTonyMasken}. For all data maps, the monopole and dipole were removed.

The three maps based on the co-added VW-band map will be treated slightly different to the analysis process from above: Instead of using the usual 20 FSS maps, we compare the VW7m, the VW7p, and the VW7pe map with 20 simulations of the VW-band, where we likewise excluded the respective mask or point sources and filled it with Gaussian noise. In doing so, we avoid to detect deviations that are only due to the structural differences inside the masked regions or to the peculiar boundary between the masks and the rest of the map. These unintentional deviations might occur when applying the FSS method. However, for the VW7m map, a comparison with FSS was tested as well, leading to similar results as the approach with the masked simulations.

The ILC7 and NILC5 data sets are maps that can also be used in a full sky surrogate analysis. The maps were nevertheless analysed in this work, on the one hand to check for consistency with the full sky surrogates method, and on the other hand to elucidate the influence of the different map making procedures on the phase correlations. Apparently, effects of these map making procedures should in particular be present in the Galactic Plane.

For the VW7m map, a full sky surrogate analysis would not be applicable, since obviously the filling of the mask has a severe influence on the phase correlations. However, it could be used in a regular full sky analysis, e.g. in comparison with simulated CMB maps of the VW band, for which the same mask filling method is applied.

The remaining two maps (VW7p and VW7pe) represent data sets that do not attempt to remove the foreground influences of the Milky Way. These maps would not produce reasonable results when analysed with the full sky surrogates approach. Thus, they are a perfect example to demonstrate the usage of the method of cut sky surrogates. By applying the method on these maps, we are consistent with the statement that the removal of foreground affected regions is to be preferred to an reconstruction of these areas \cite{Copi11}.

\subsection{Results}

\begin{figure}
\centering
\includegraphics[width=4.27cm, keepaspectratio=true, ]{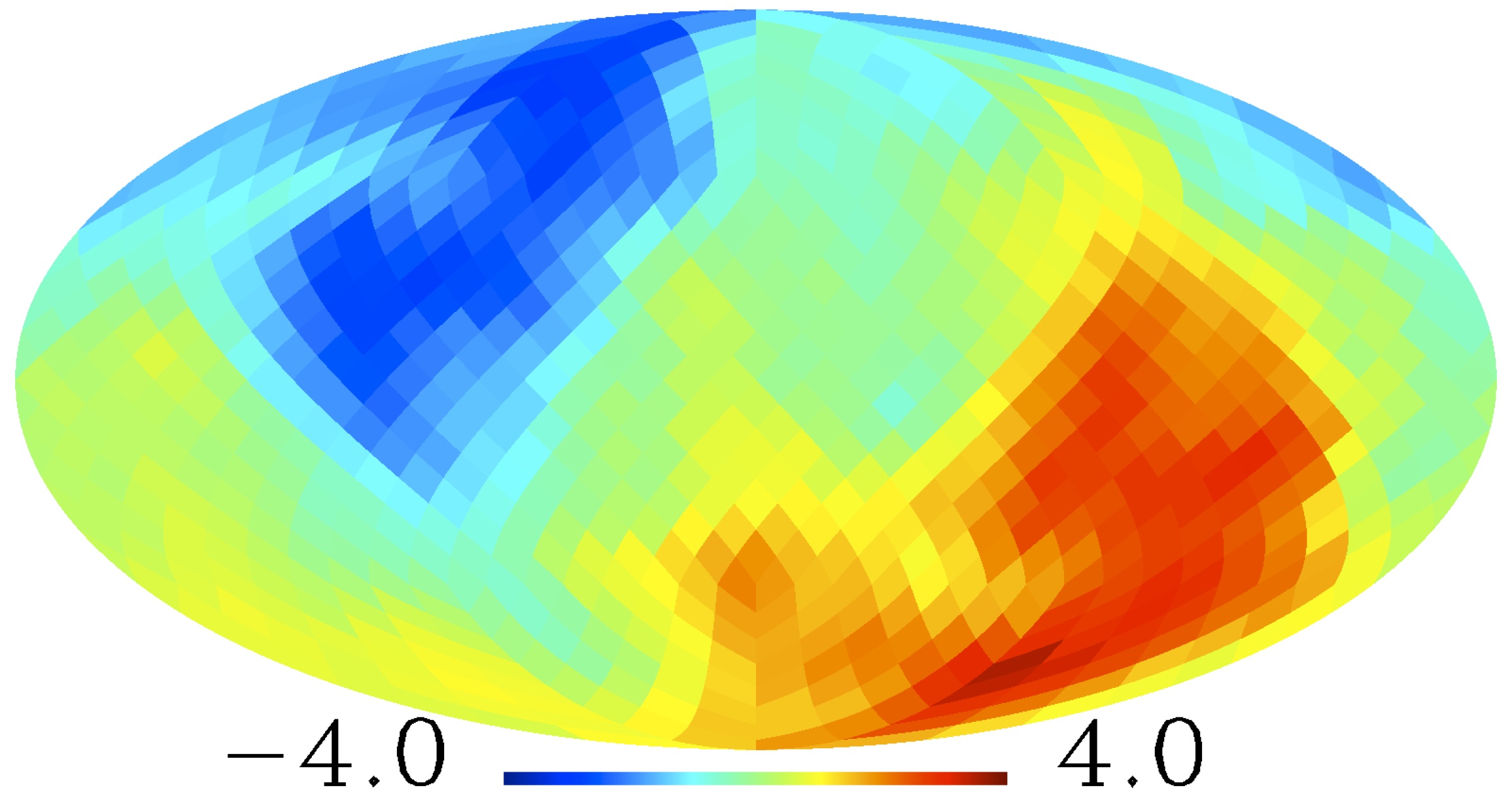}
\includegraphics[width=4.27cm, keepaspectratio=true, ]{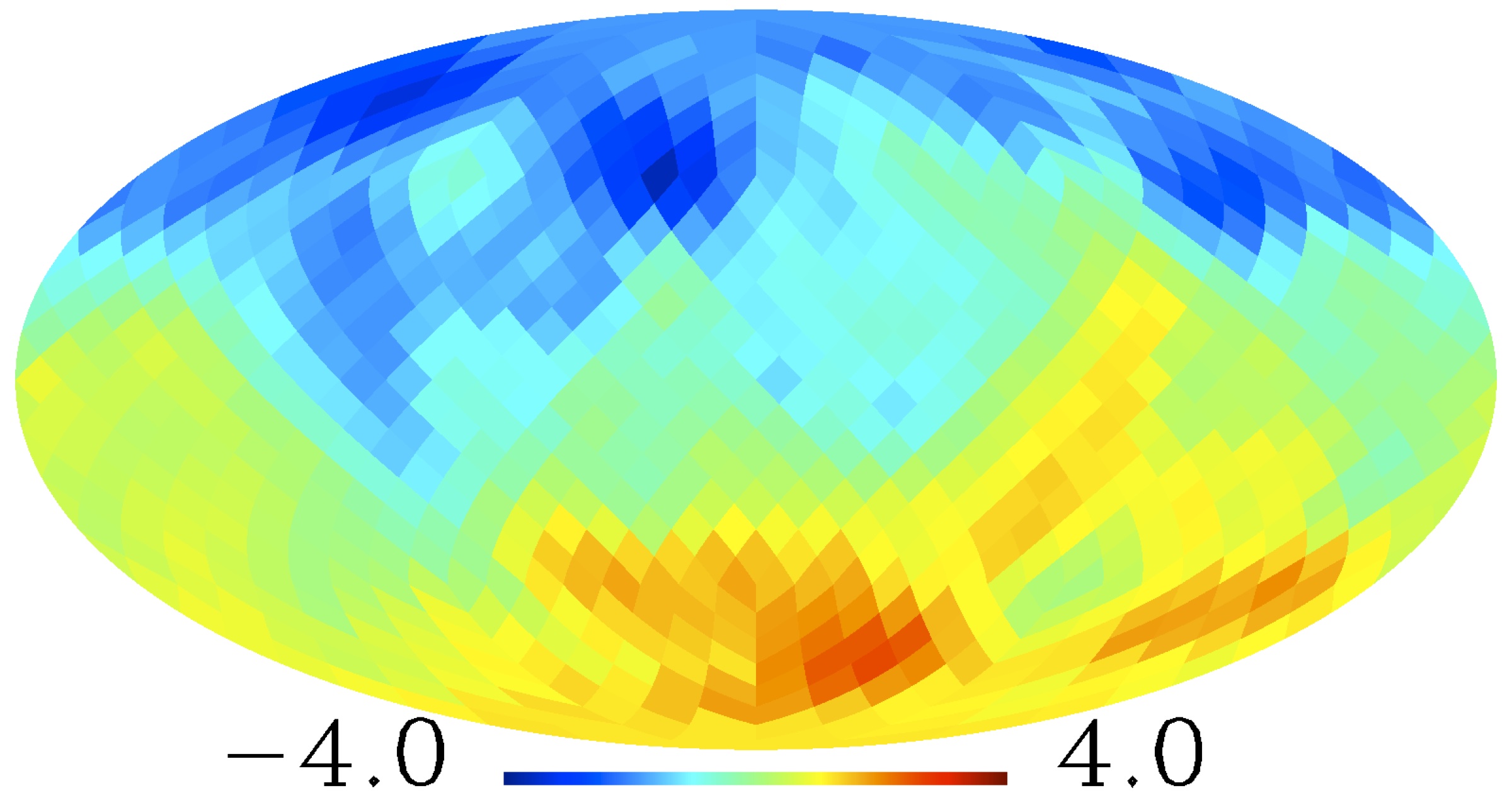}
\includegraphics[width=4.27cm, keepaspectratio=true, ]{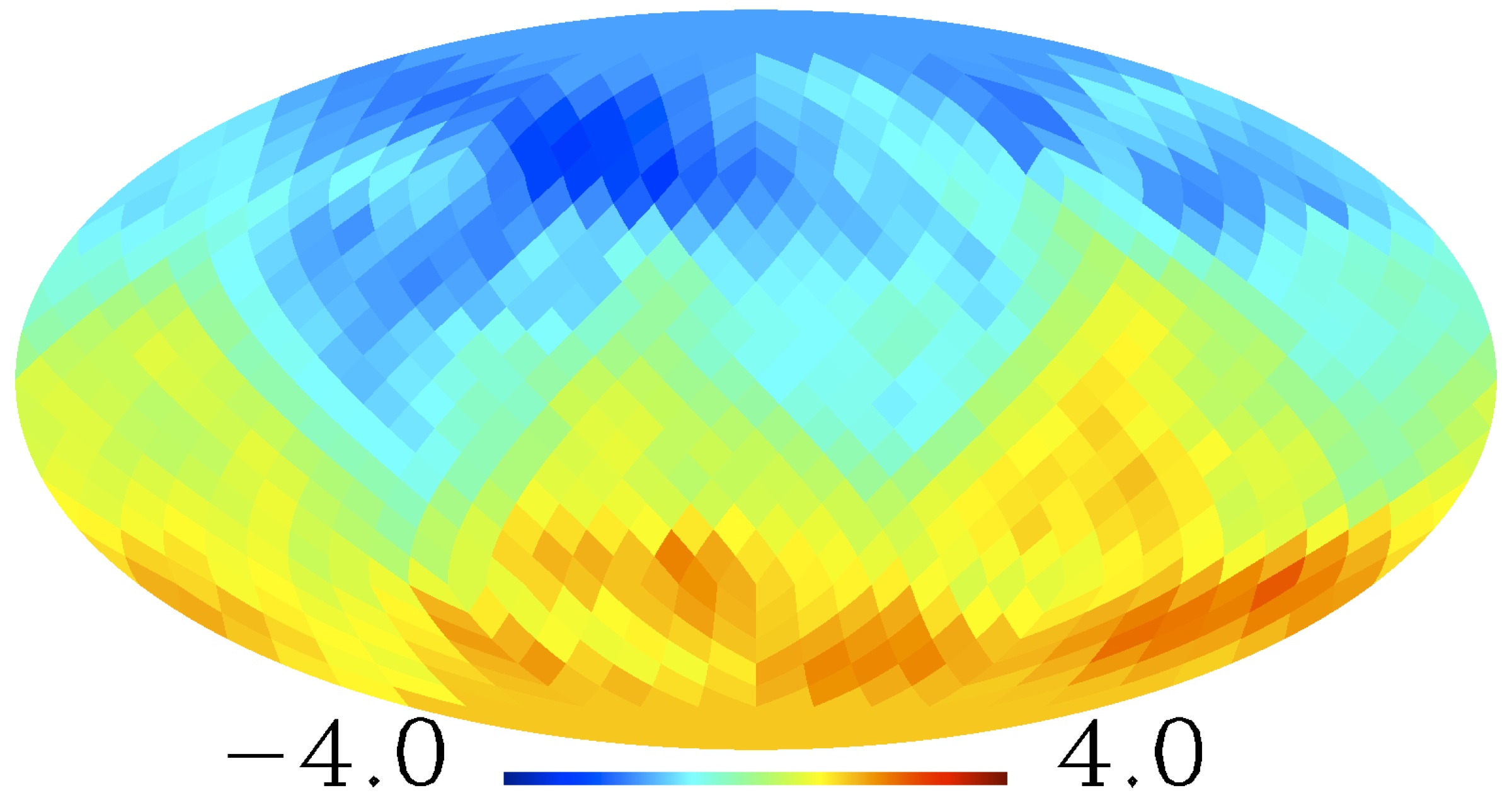}
\includegraphics[width=4.27cm, keepaspectratio=true, ]{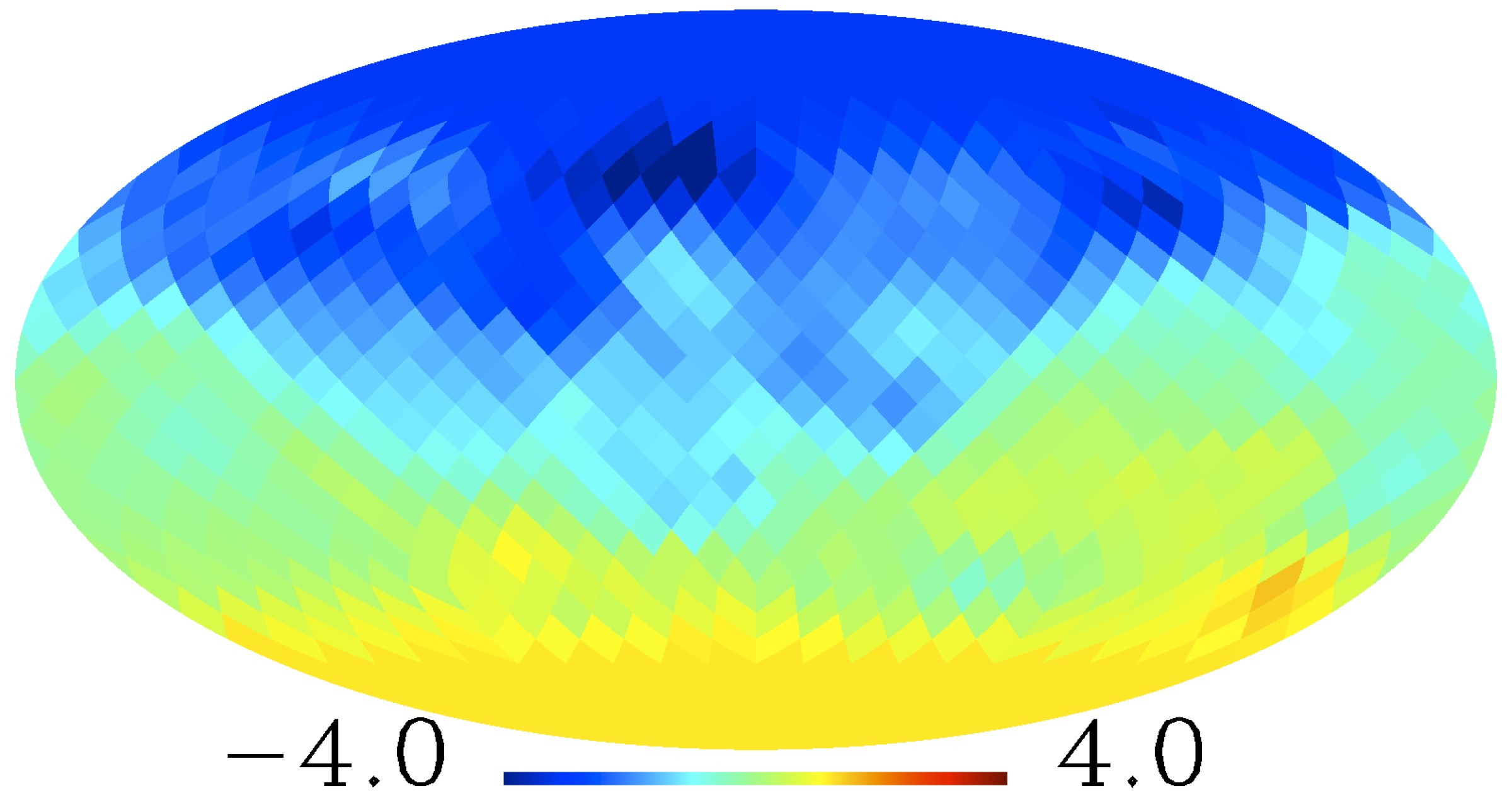}
\caption{Same as figure \ref{Heikeplots} for the NILC5 map.} \label{Mollplots}
\end{figure}

\begin{figure}
\centering
\includegraphics[width=4.27cm, keepaspectratio=true, ]{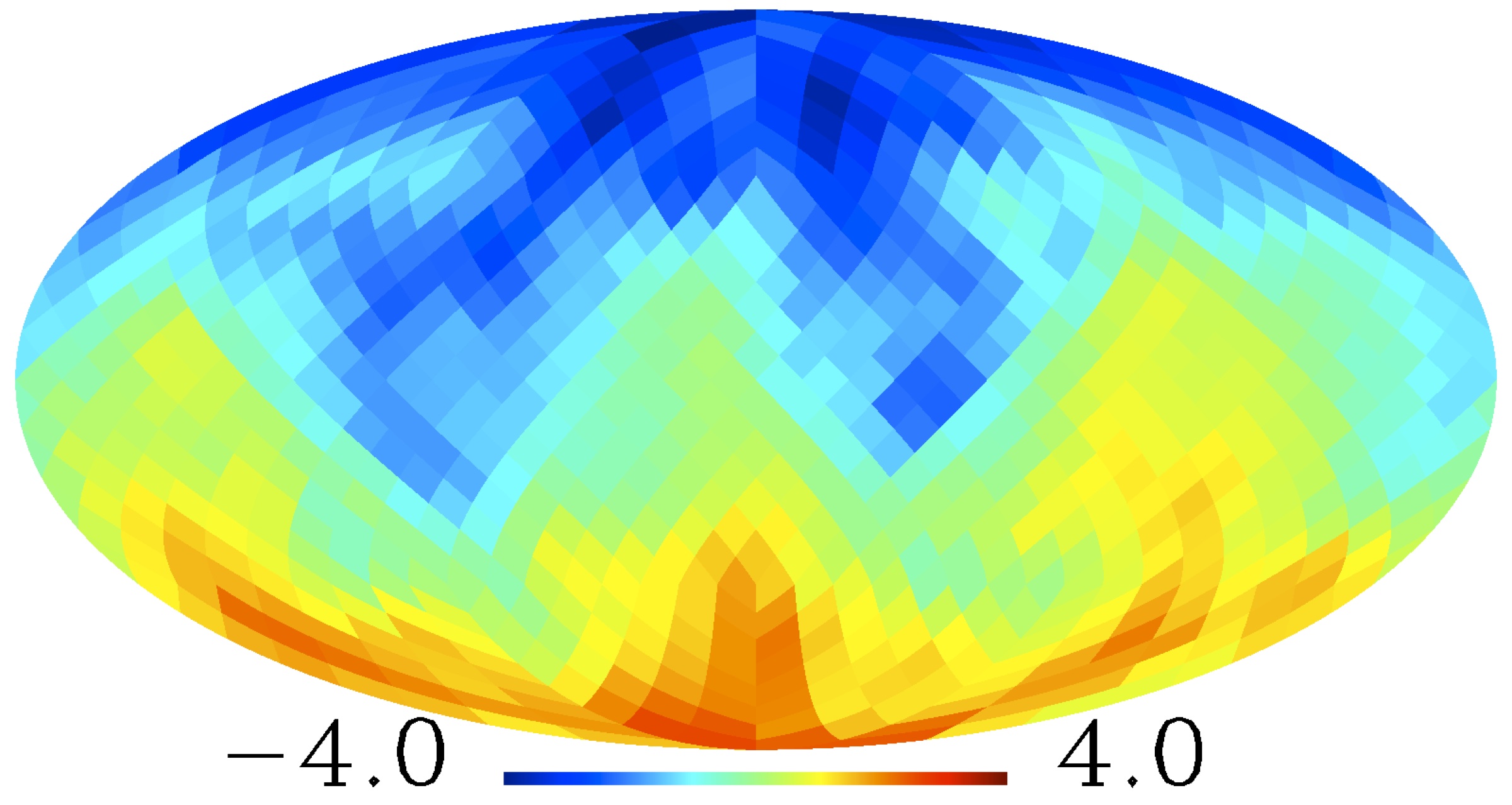}
\includegraphics[width=4.27cm, keepaspectratio=true, ]{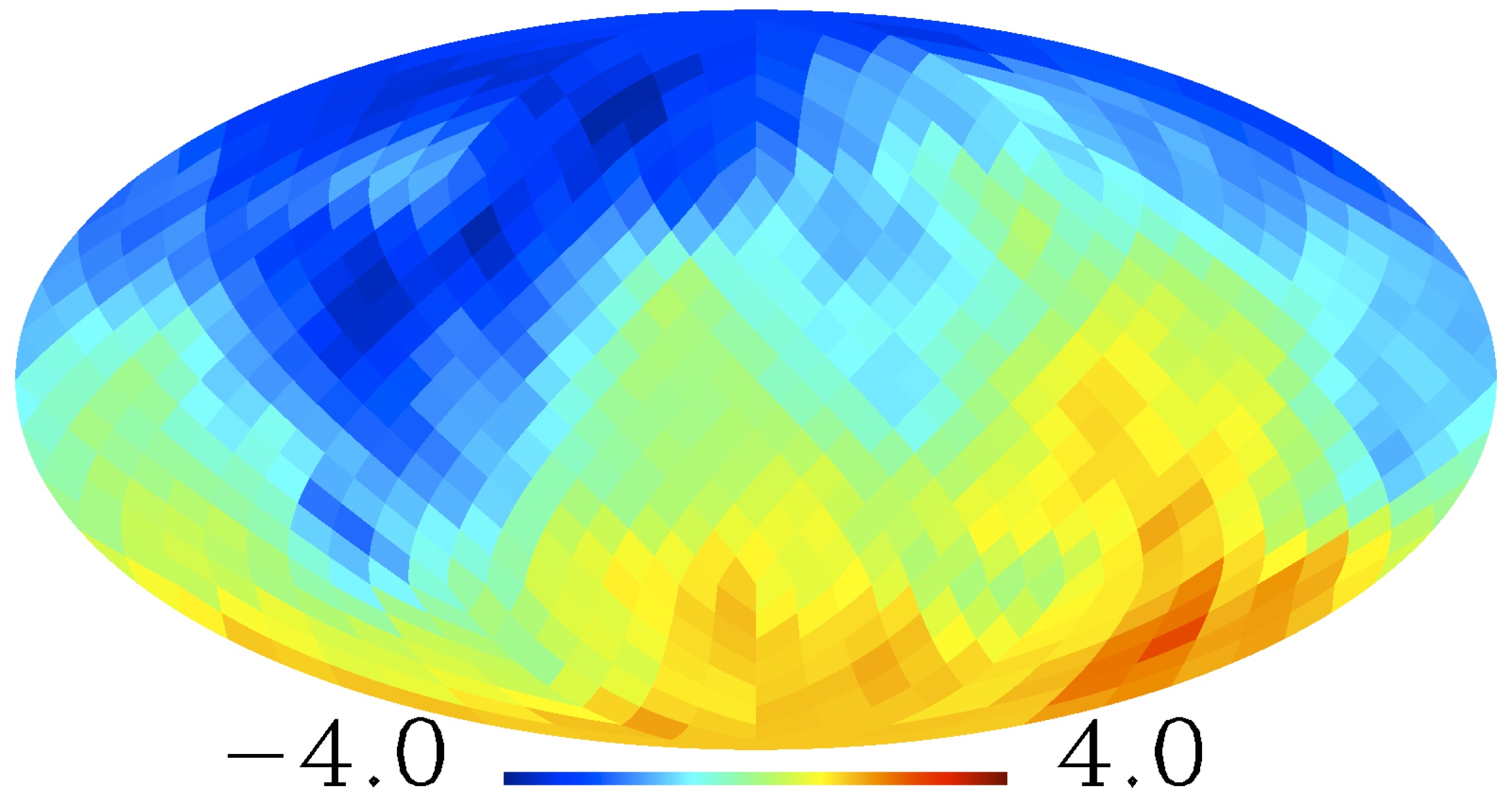}
\includegraphics[width=4.27cm, keepaspectratio=true, ]{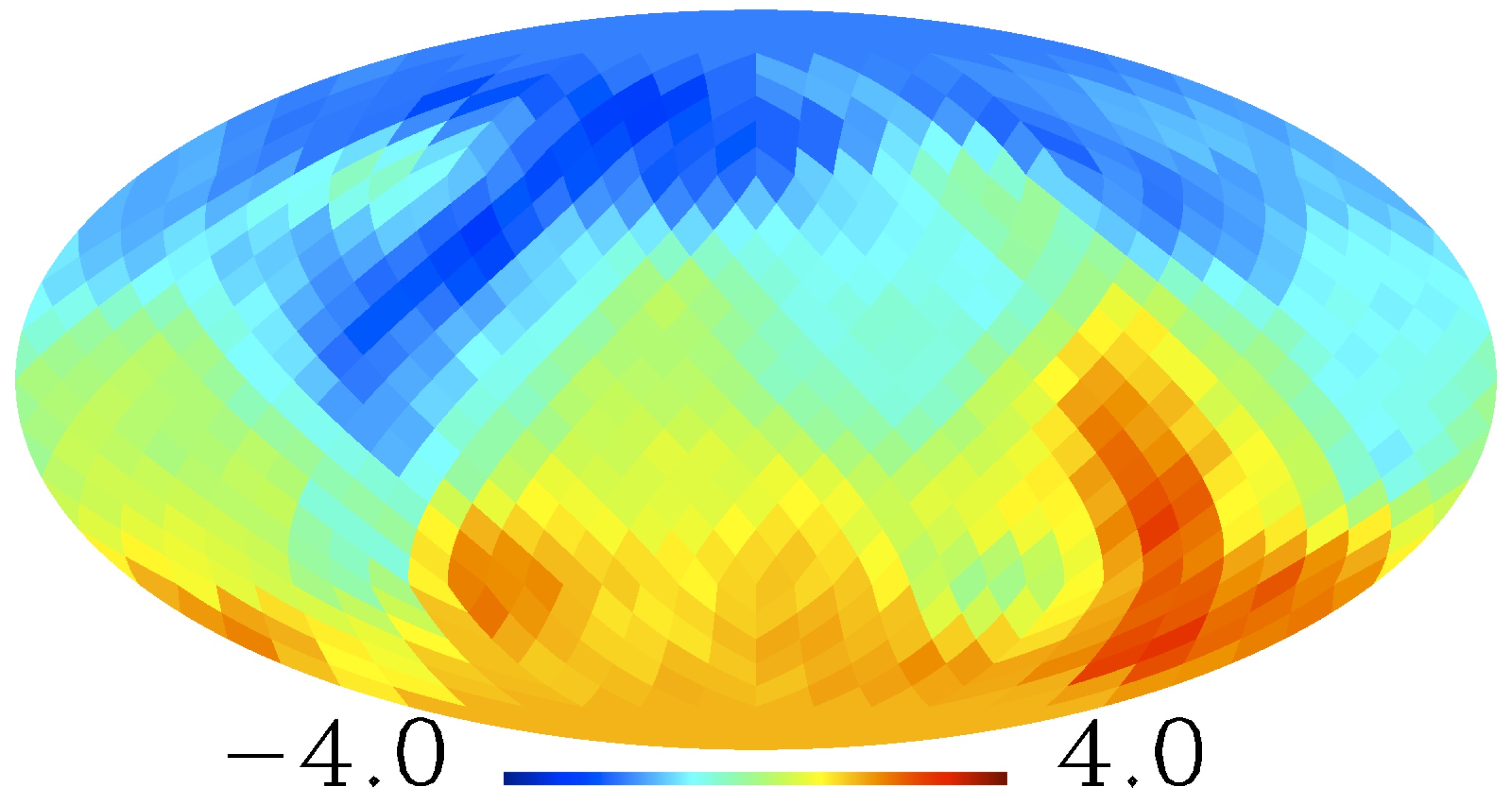}
\includegraphics[width=4.27cm, keepaspectratio=true, ]{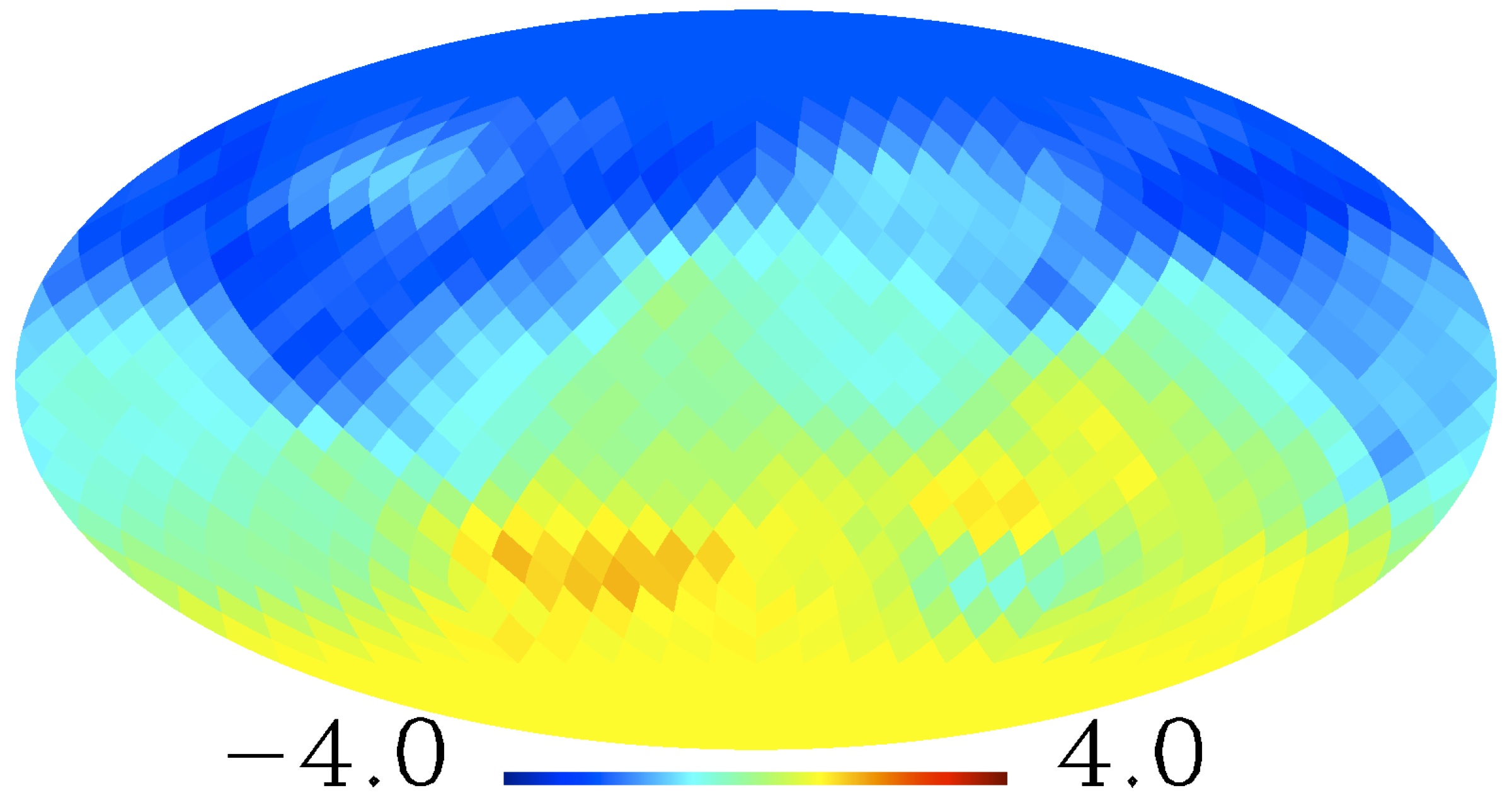}
\caption{Same as figure \ref{Heikeplots} for the VW7pe map.} \label{VWcoadd7PointsExt1}
\end{figure}

When evaluating the scaling index results $S_2(\langle \alpha \rangle)$ and $S_2(\sigma_\alpha)$ for the $|b| <  10^\circ$ cut for the rotated hemispheres of all data maps described above, we detect significant non-Gaussianities and an asymmetry. Both features were already found in corresponding full-sky analyses \cite{Raeth09,Raeth10}. The signal for $S_2(\langle \alpha \rangle)$ becomes less clear for larger cuts. But for $S_2(\sigma_\alpha)$, a significant asymmetry with a clear north-south direction persists for all maps when excluding the Galactic plane. As examples, the significances $S_2(\sigma_\alpha)$ determined for the rotated hemispheres of the NILC5 and the VW7pe map for the different sky cuts are shown in the figures \ref{Mollplots} and \ref{VWcoadd7PointsExt1}, respectively. 

\begin{figure}
\centering
\includegraphics[width=4.27cm, keepaspectratio=true, ]{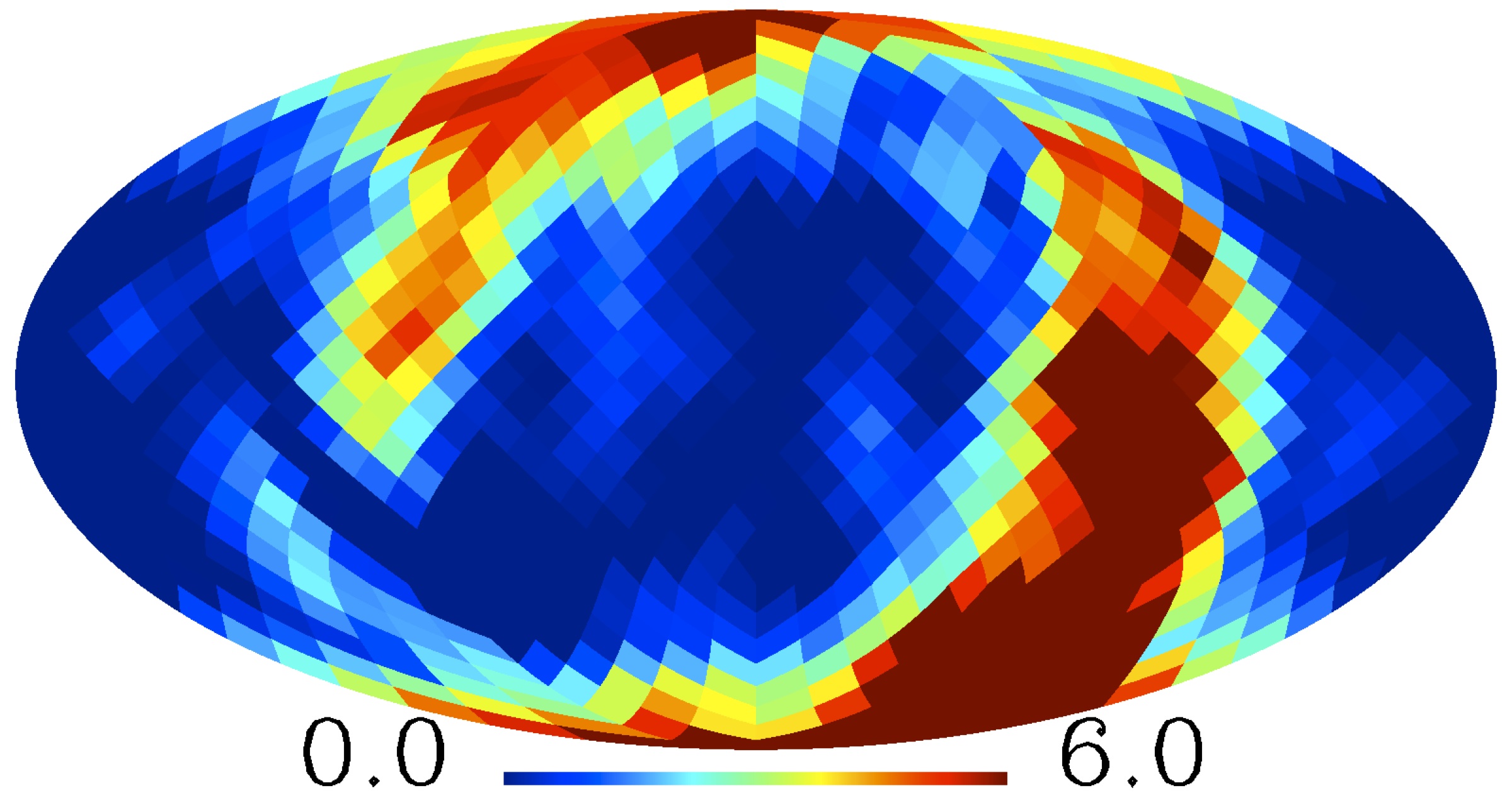}
\includegraphics[width=4.27cm, keepaspectratio=true, ]{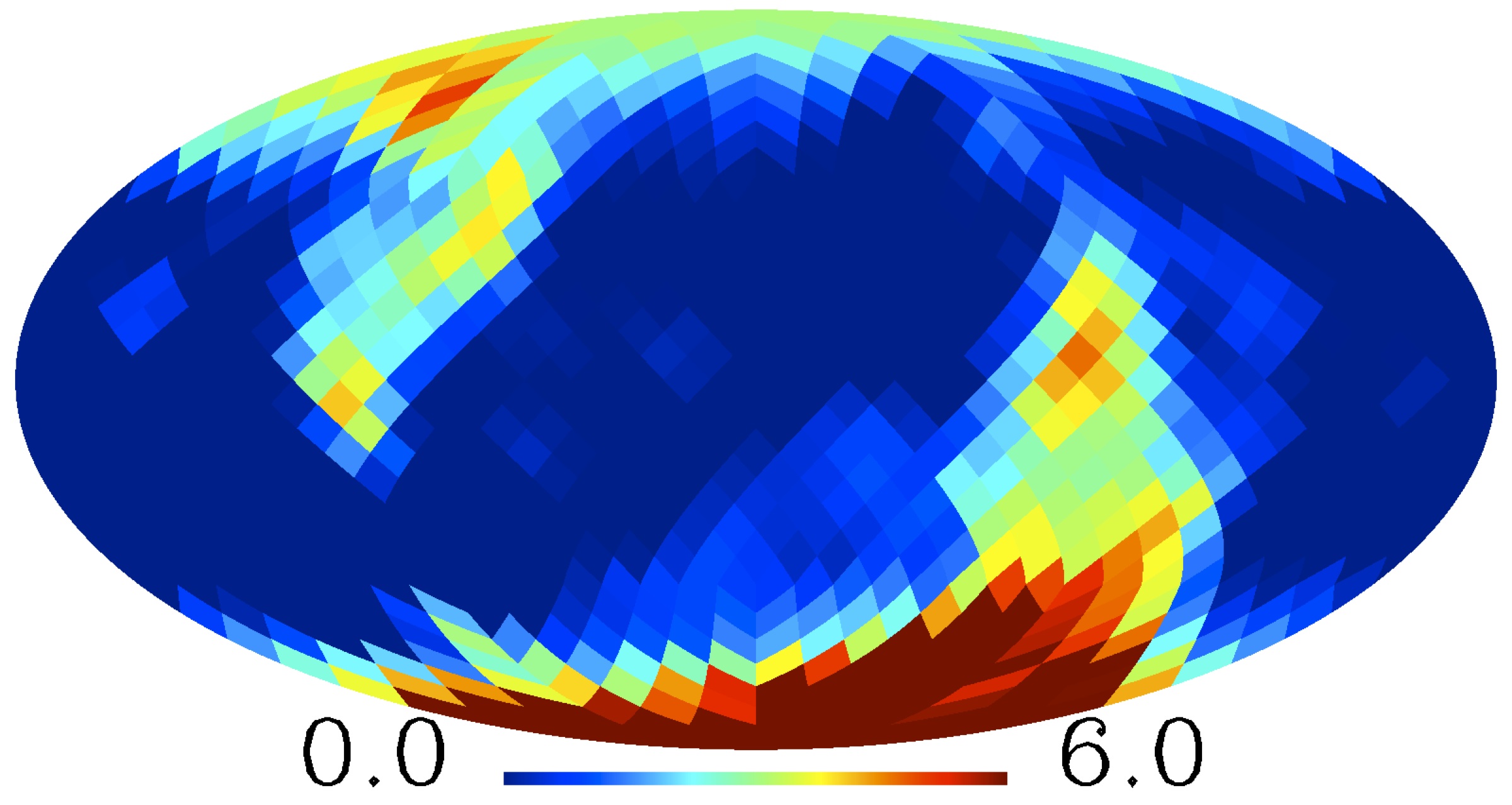}
\includegraphics[width=4.27cm, keepaspectratio=true, ]{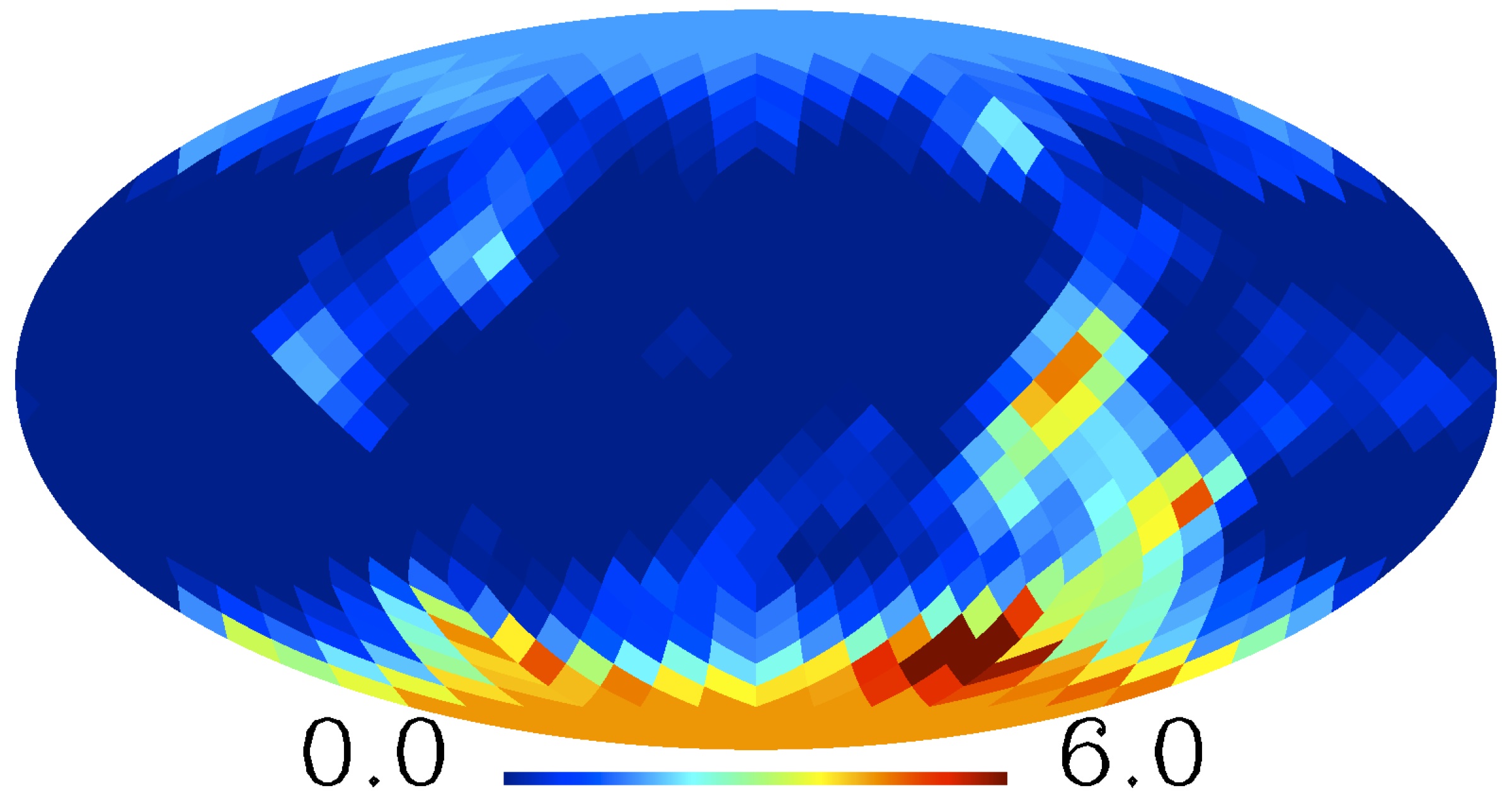}
\includegraphics[width=4.27cm, keepaspectratio=true, ]{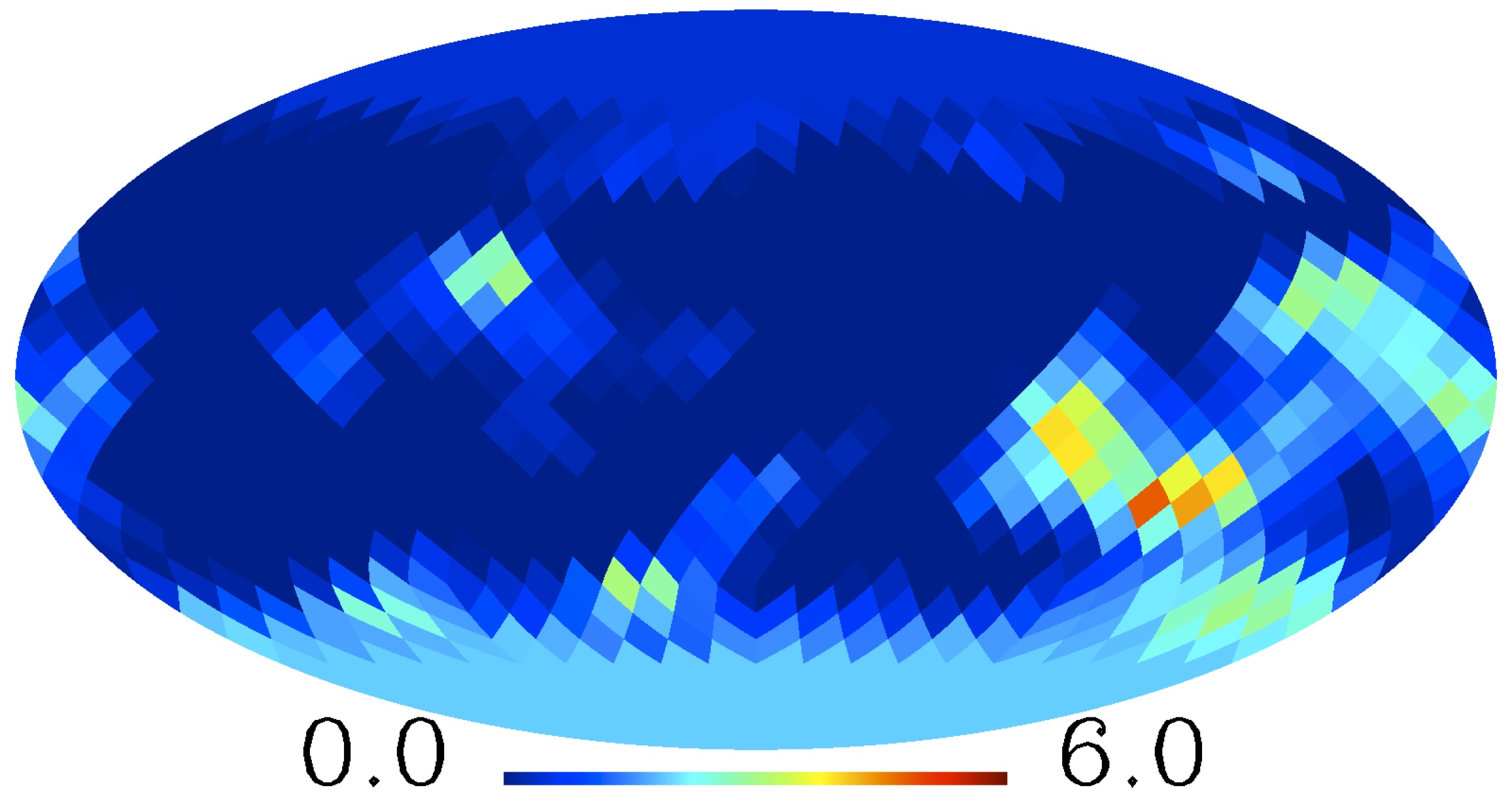}
\caption{Same as figure \ref{Heikeplots2} for the NILC5 map.} \label{EchteHeikeplots}
\end{figure}

\begin{figure}
\centering
\includegraphics[width=4.27cm, keepaspectratio=true, ]{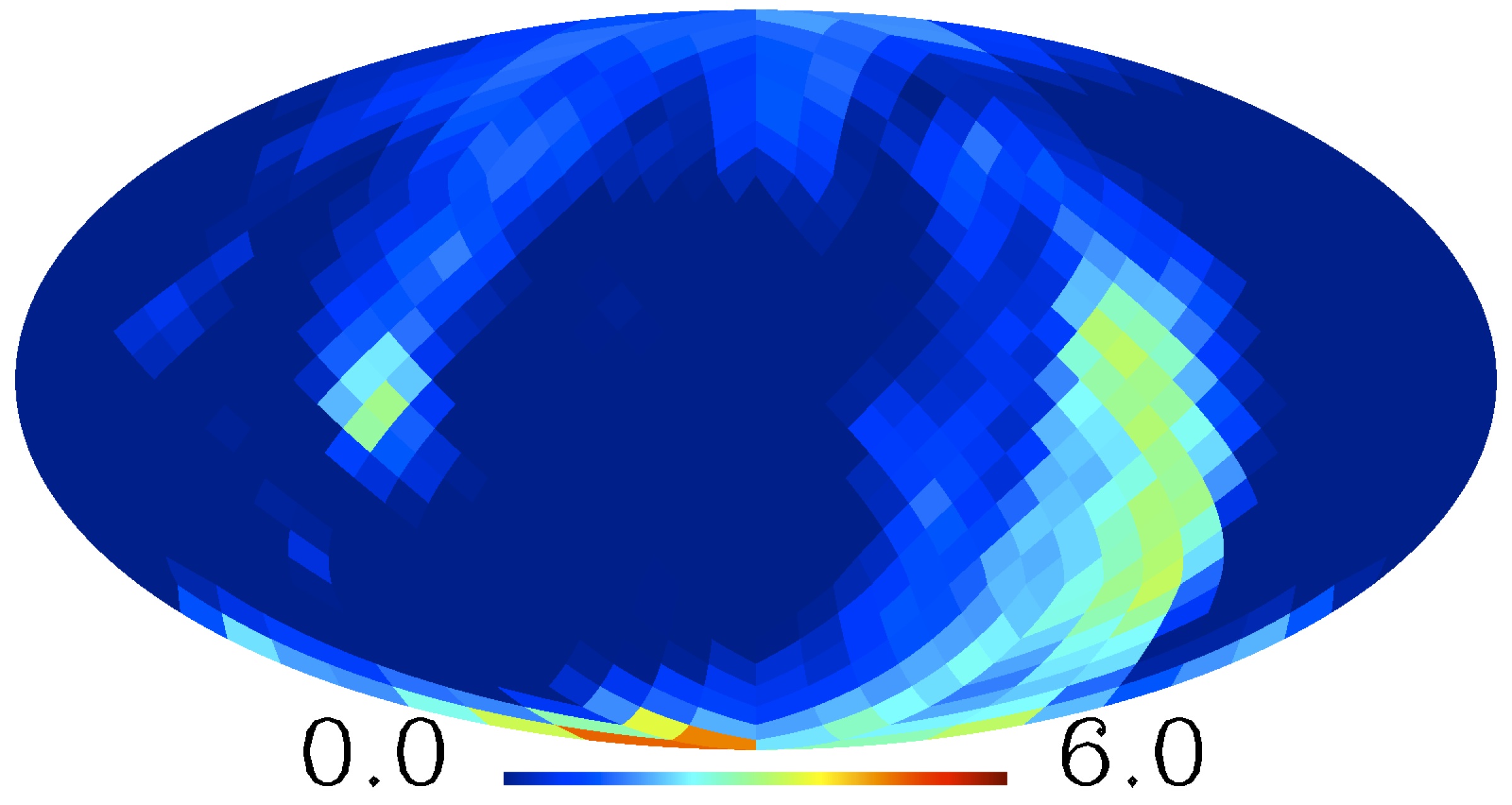}
\includegraphics[width=4.27cm, keepaspectratio=true, ]{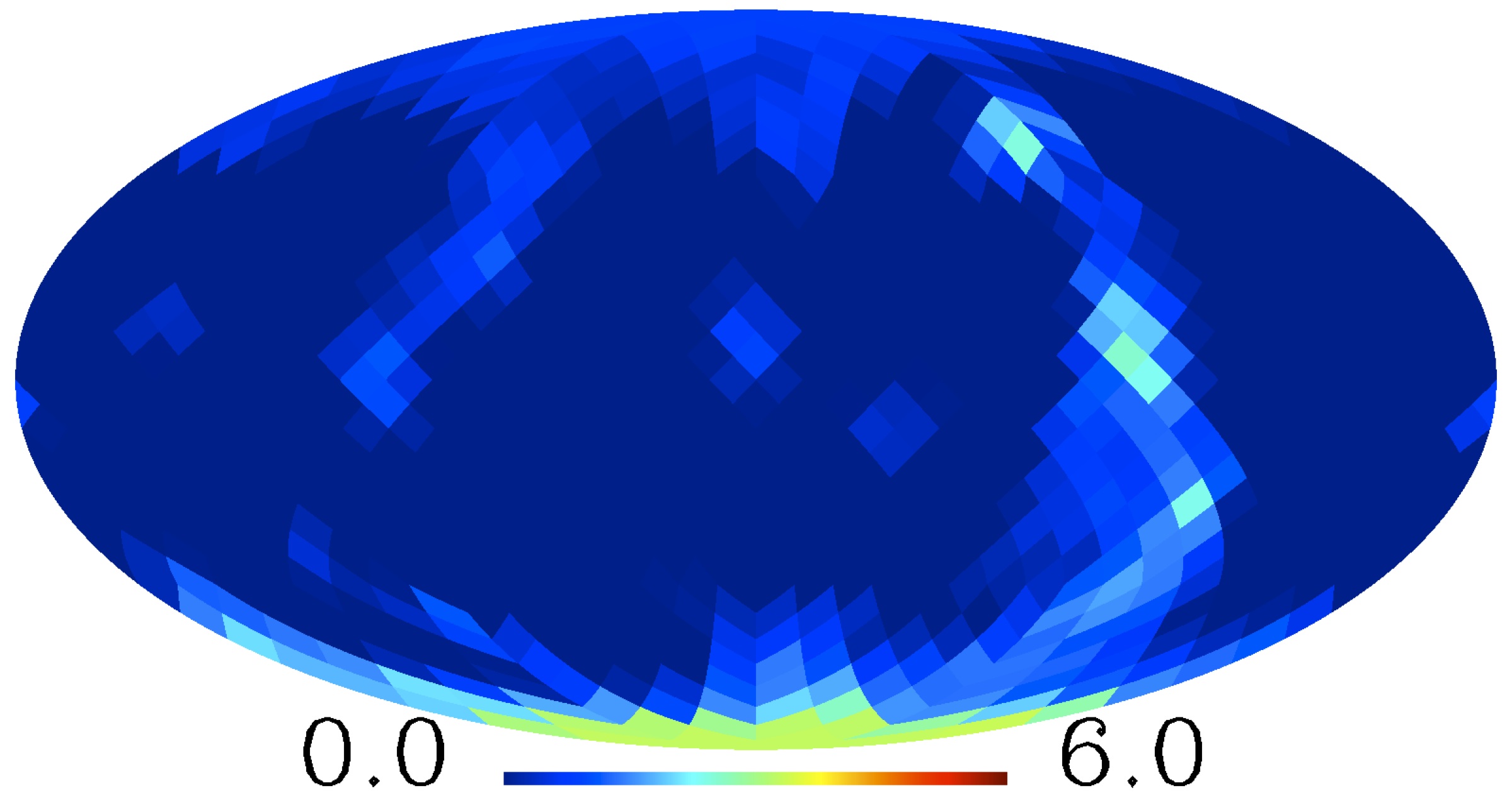}
\includegraphics[width=4.27cm, keepaspectratio=true, ]{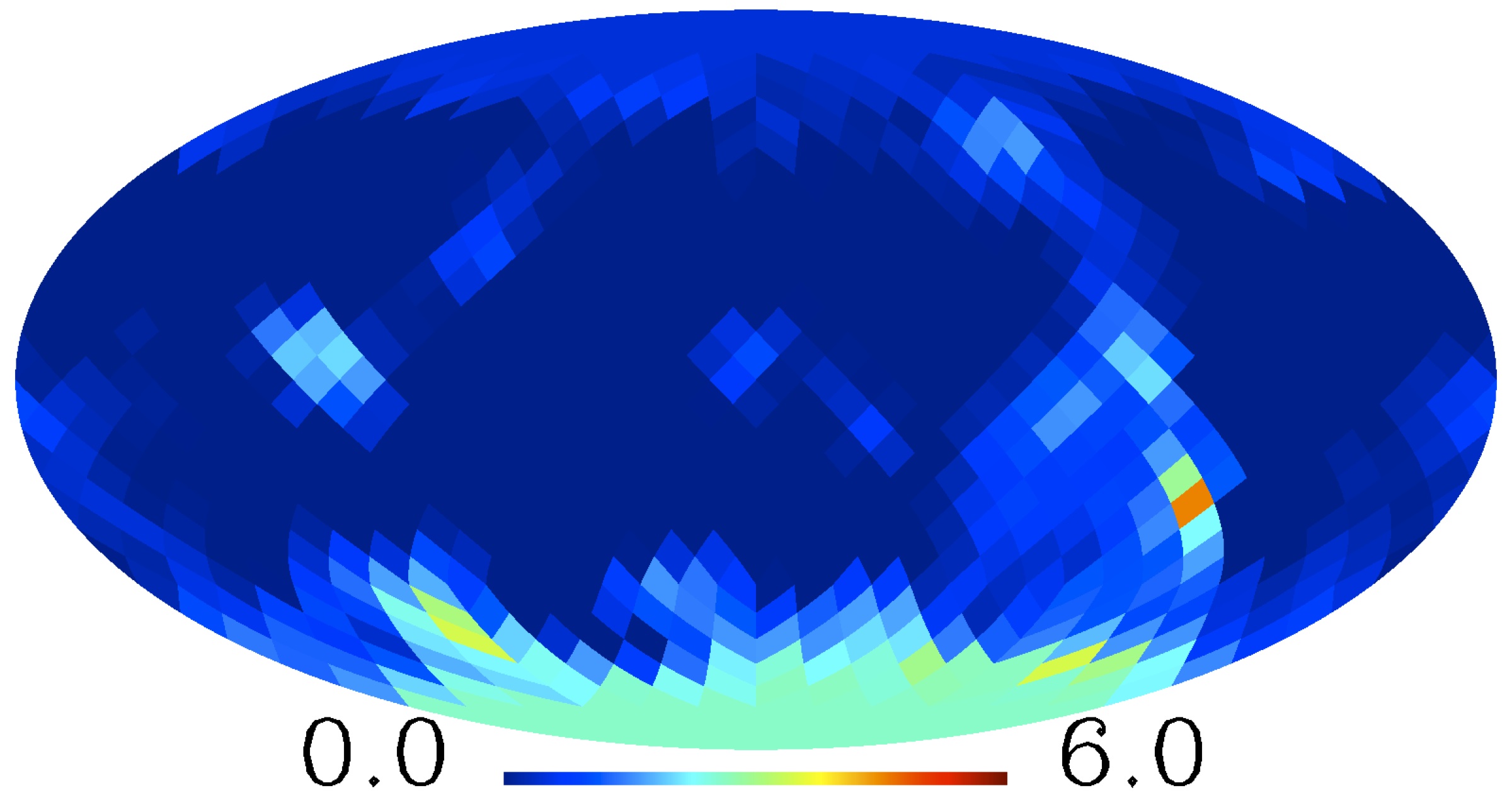}
\includegraphics[width=4.27cm, keepaspectratio=true, ]{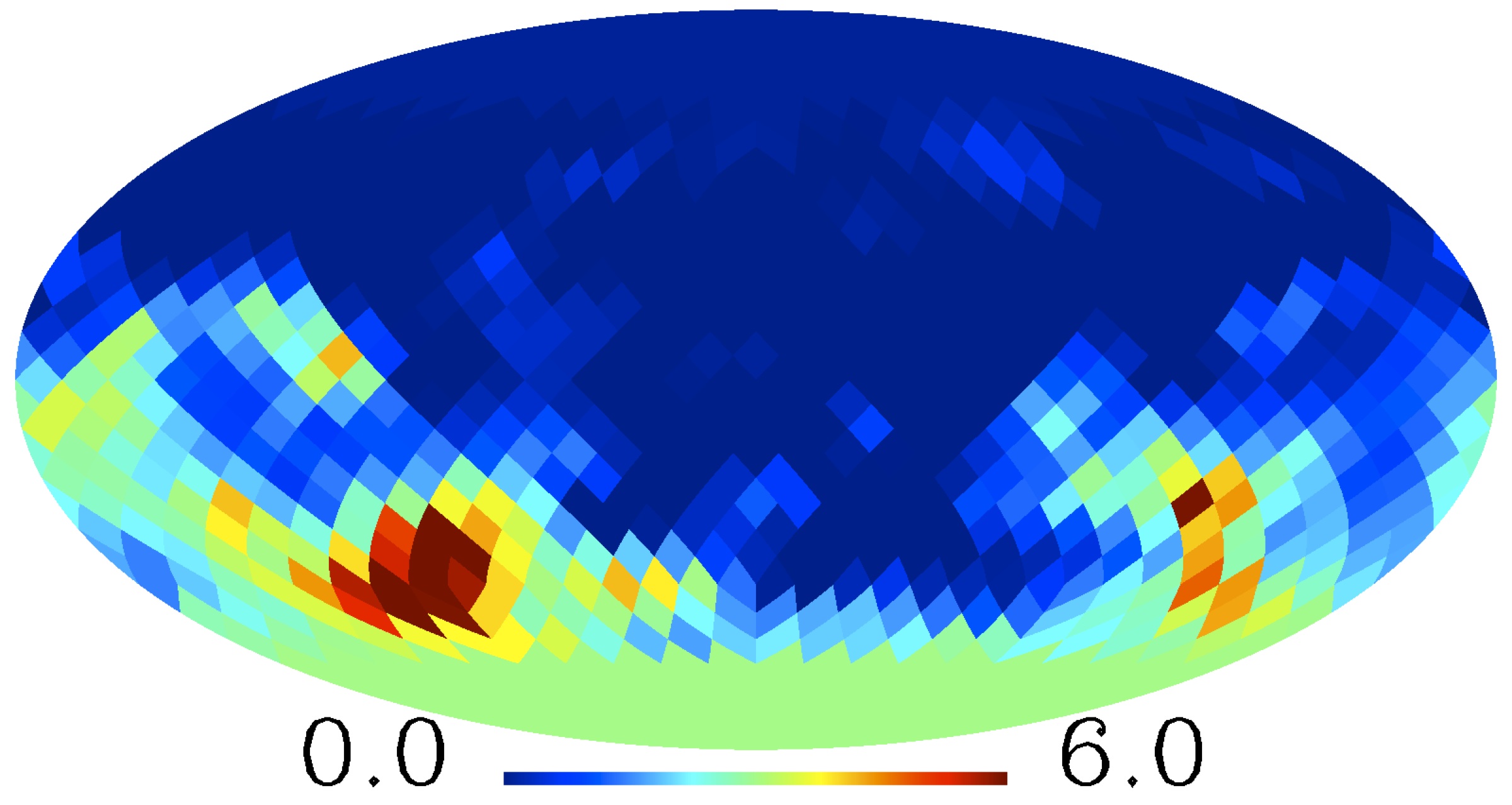}
\caption{Same as figure \ref{Heikeplots2} for the VW7pe map.} \label{VWcoadd7PointsExt2}
\end{figure}

The Minkowski functionals show similar results: For $M_{area}(\nu)$, one detects similar deviations between all the data sets and its related full sky surrogates. The only exception is the VW7pe map, where the asymmetries are less pronounced than in the other maps. For the VW7m map, that cuts out even larger areas of foreground affected data, a clear asymmetric behaviour is found again. As examples, the results for $M_{area}(\nu)$ are illustrated for the NILC5 and VW7pe maps in the figures \ref{EchteHeikeplots} and \ref{VWcoadd7PointsExt2}. 

For $M_{perim}(\nu)$ and $M_{euler}(\nu)$, the results agree for the full sky, but get less definite for larger cuts. However, experience shows that this could likely be due to the limited amount of pixels that is present especially when applying the $|b| < 20^\circ$ or $30^\circ$ cuts. The decreasing number of pixels on the incomplete skies leads to an increasing influence of noise and a lower influence of the intrinsic signal. Due to discreetness, possible effects on the tails of the Minkowski functionals may vanish, which in turn might damp the respective $\chi^2$. This especially holds for the two Minkowski functionals $M_{perim}(\nu)$ and $M_{euler}(\nu)$ that examine complex pixel formations and thus need enough data points to produce statistically reliable results.

\begin{figure}
\centering
\includegraphics[width=8.5cm, keepaspectratio=true, ]{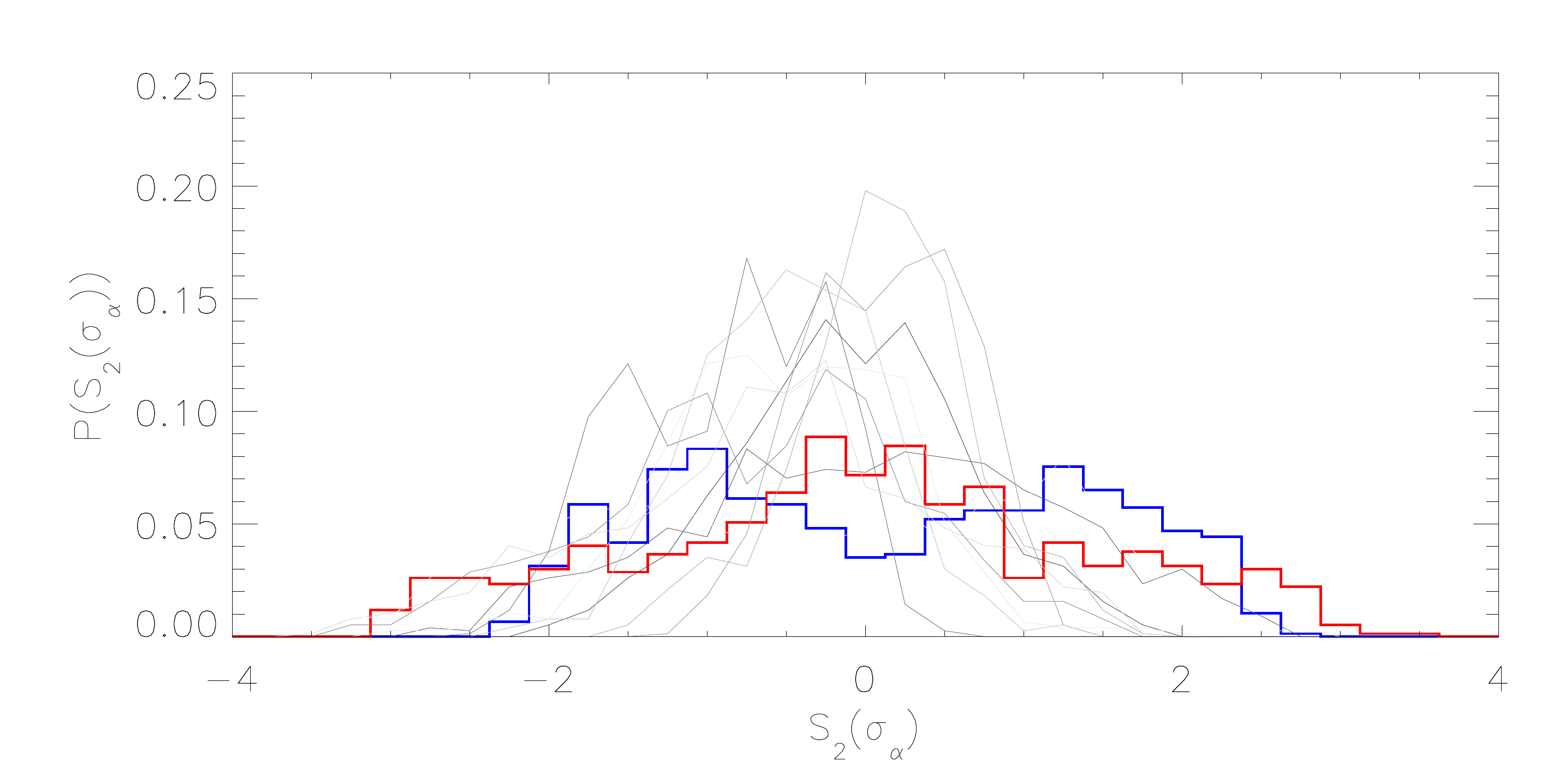}
\includegraphics[width=8.5cm, keepaspectratio=true, ]{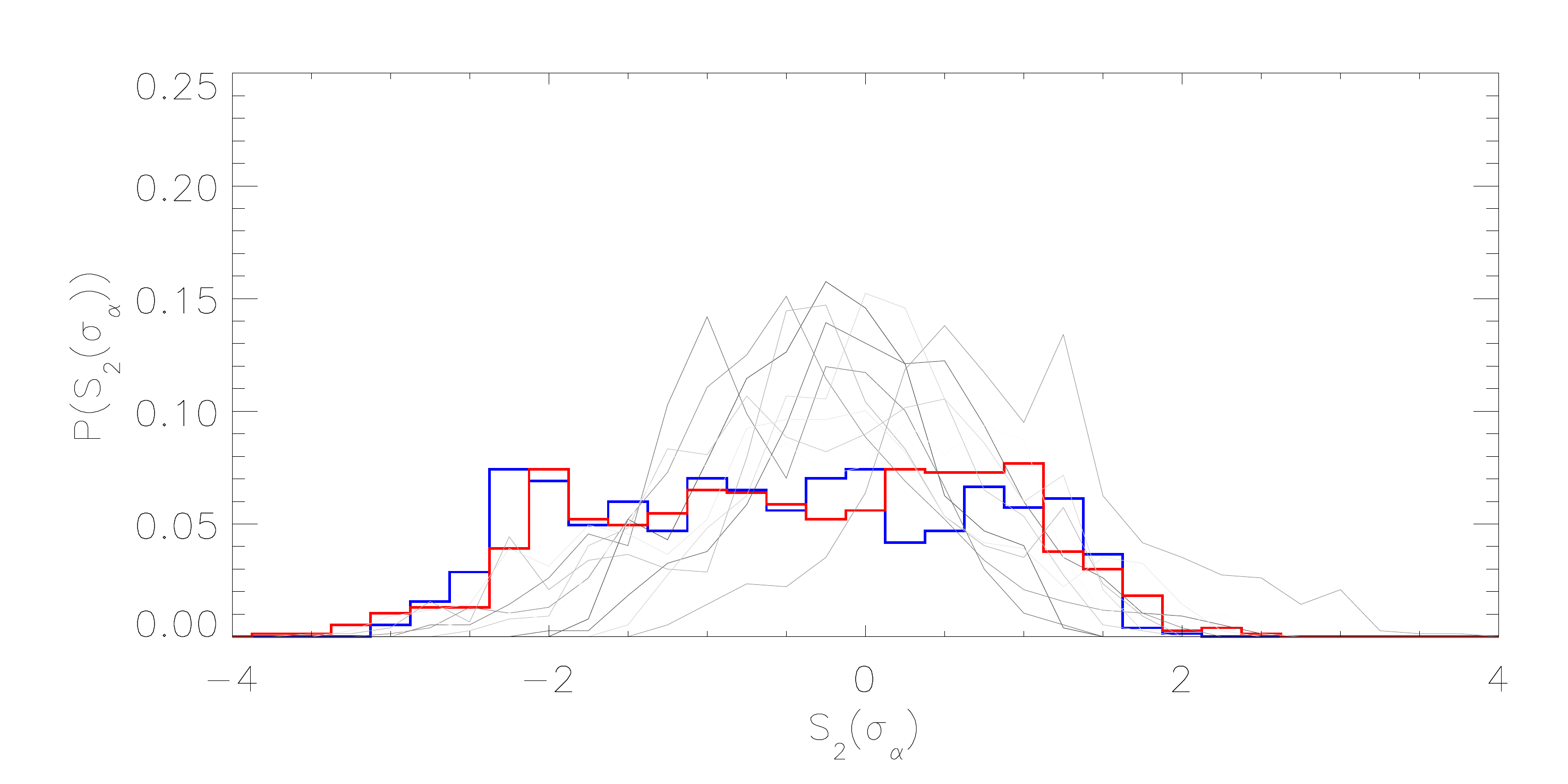}
\includegraphics[width=8.5cm, keepaspectratio=true, ]{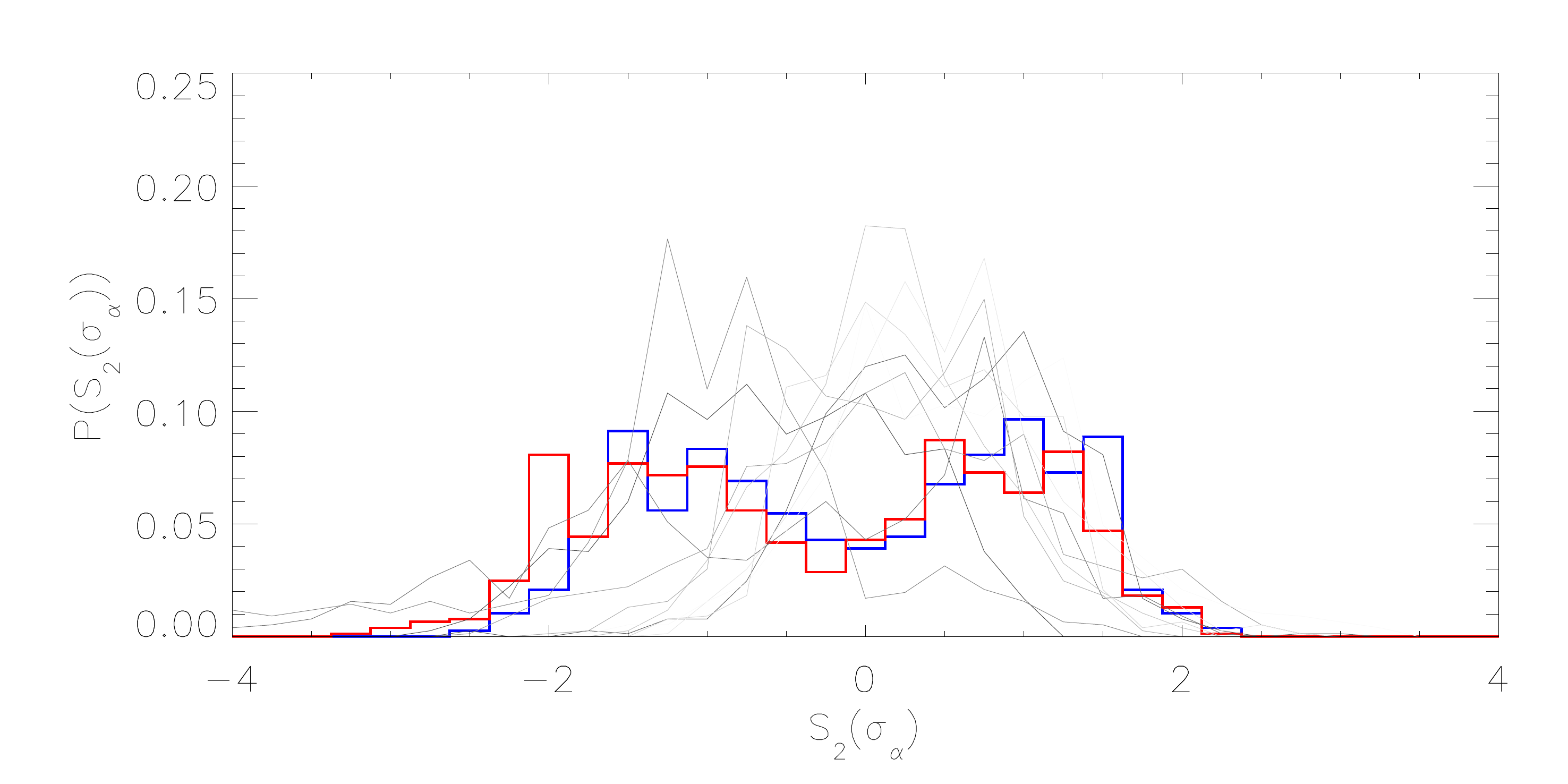}
\includegraphics[width=8.5cm, keepaspectratio=true, ]{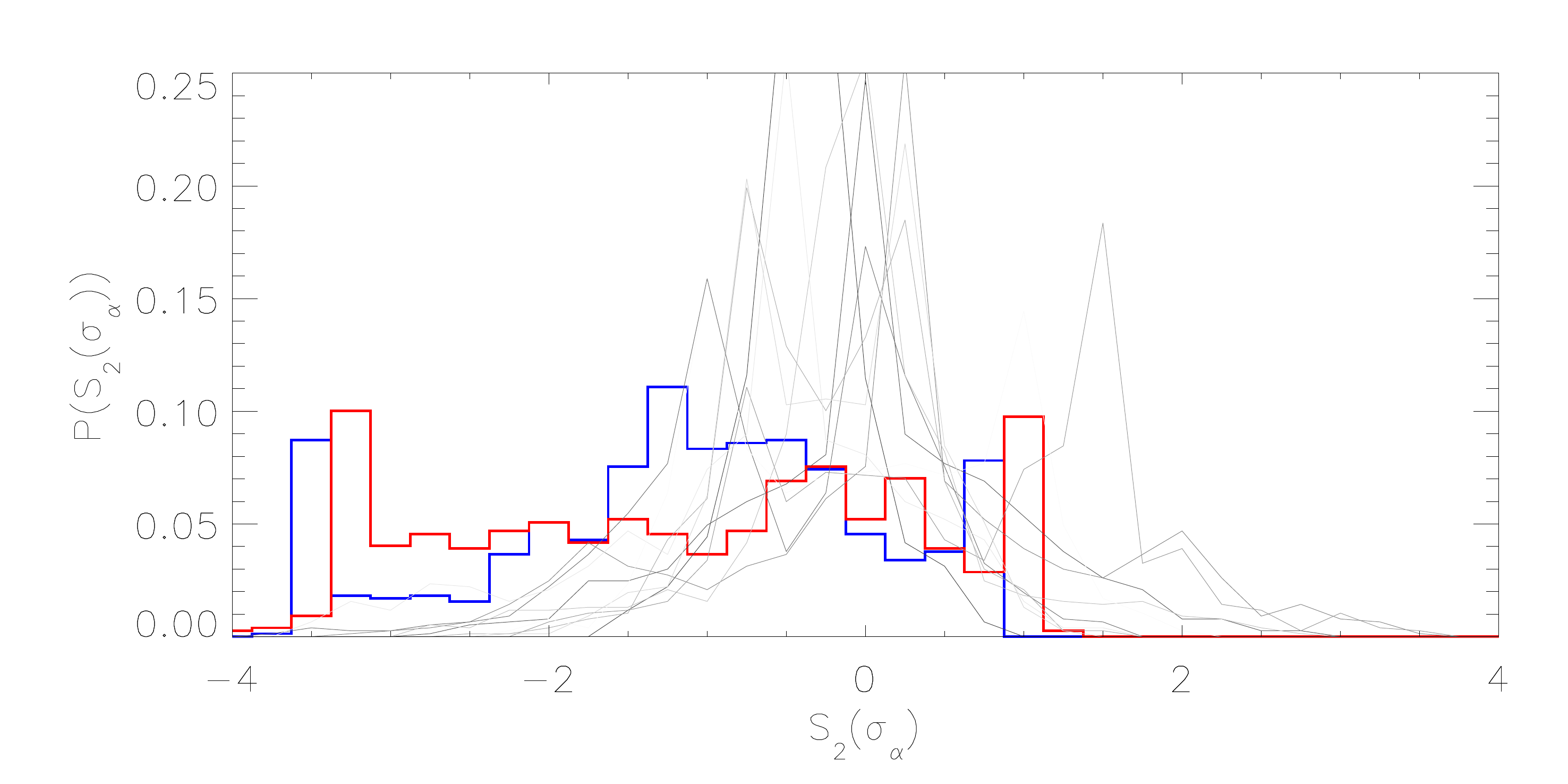}
\caption{The histograms of $S_2(\sigma_{\alpha})$ for the complete sphere and the three different central latitude sky cuts $|b| < 10^\circ$, $|b| < 20^\circ$, and $|b| < 30^\circ$ (from top to bottom), for the ILC7 (blue) and NILC5 (red) maps as well as for 10 simulated Gaussian random fields without noise (grey).} \label{Histograms}
\end{figure}

The detected north-south asymmetry is reflected in the empirical distribution functions of the result maps $S_2(\sigma_\alpha)$. We compare the results of the ILC7 and the NILC5 maps with 10 simulations without noise. These simulated maps undergo the same processes as the data maps. That means we constructed for each simulation 20 FSS, and again for each FSS and the original map 100 CSS for all different sky cuts. The results are illustrated in figure \ref{Histograms}. Clearly, the histograms of the data maps show a much broader spectrum than the respective output of the simulations, which underlines the detection of statistical anisotropies.

\begin{figure}
\centering
\includegraphics[width=8.5cm, keepaspectratio=true]{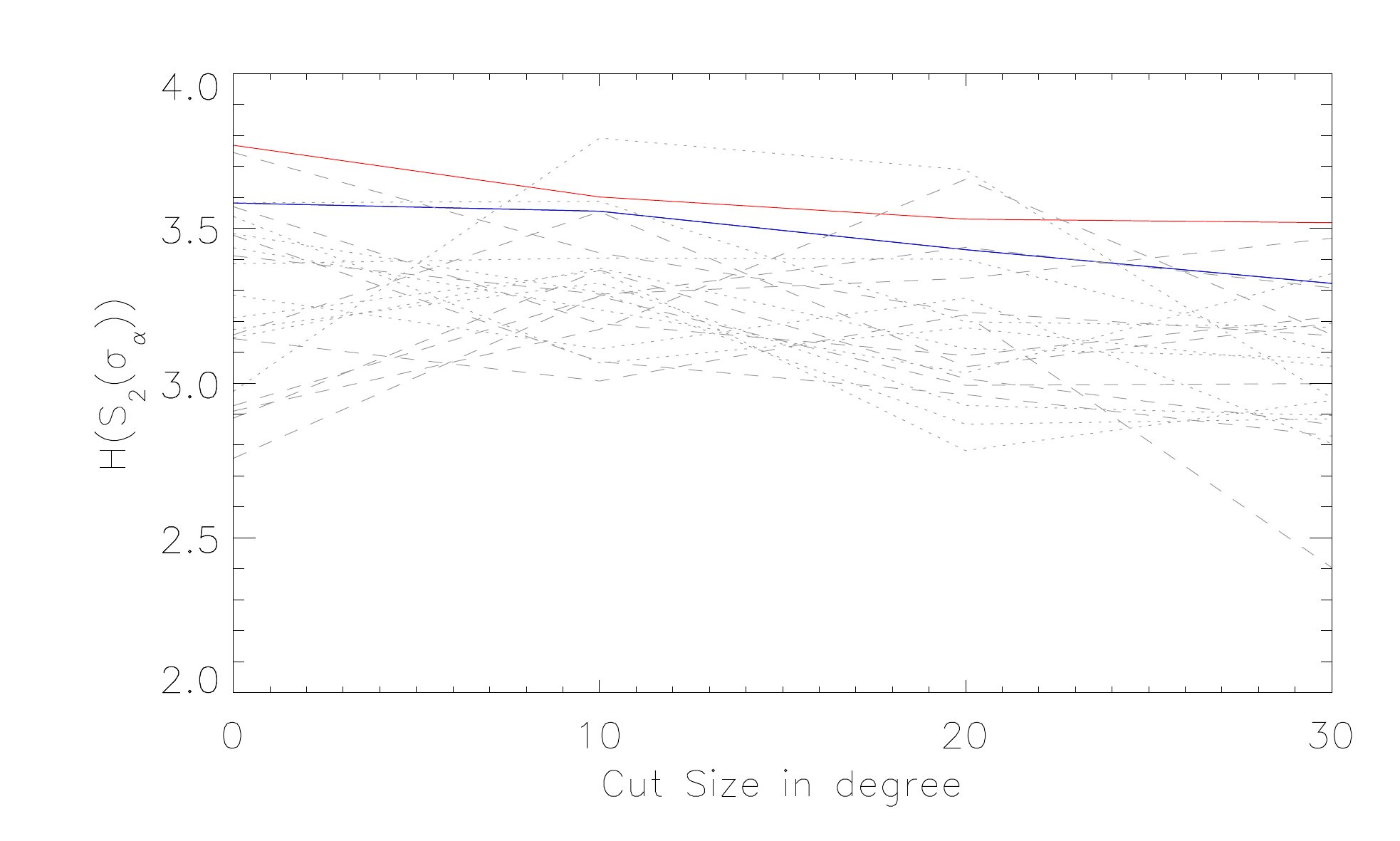}
\caption{The empirical discrete entropy $H(S_2(\sigma_{\alpha}))$ corresponding to the histograms of figure \ref{Histograms}. The blue and red lines represent the ILC7 and NILC5 maps, while the results of the simulated Gaussian random fields with and without noise (10 realisations each) are illustrated by the dotted and dashed grey lines.} \label{AsymmPlot}
\end{figure}

This aspect is in turn quantified by the empirical entropy $H(S_2(\sigma_\alpha))$. Figure \ref{AsymmPlot} presents the corresponding results of the twelve above mentioned histograms and additional ten simulations with Gaussian noise. The results of the data maps offer higher values than the ones of the simulations. Also, it is remarkable that these values are nearly not decreasing when moving to a larger sky cut. As above in section \ref{ValidationResults}, the differences between the two types of simulations are negligible.

In summary, both the scaling indices and Minkowski functionals detect higher order correlations and thus non-Gaussianities, as well as a clear north-south asymmetry in all data sets even when large parts of the Galactic Plane are removed. Only the output of the Minkowski functionals of the VW7pe map shows a weaker significance. Nevertheless, the scaling indices detect an asymmetry in the VW7pe map. To investigate this particularity a bit further, we repeated the CSS analysis of the NILC5 map where we removed the masked regions of the VW7pe map. For simplicity reasons, no new FSS were generated. A comparison of this new NILC5pe map with the FSS of the NILC5 data lead to the same results as the analysis of the NILC5 map itself.

The different signs in the opposite hemispheres of the findings for $S_2(\sigma_\alpha)$ that appear for all data maps reflect a lower (higher) variability of the scaling index results for the northern (southern) hemisphere of the original maps compared to their surrogates. A higher variability implies the presence of different types of structure. Thus, the lower variability in the north can also be interpreted as a lack of power in the northern hemisphere, which is in agreement with former results \cite{Eriksen04b, Hansen09, Rossmanith09, Raeth09, Raeth10, Eriksen04a, Hansen04}. The regions with large deviations from isotropy become very broad for the larger cuts, which makes it difficult to draw a clear statement about a more accurate direction of the detected asymmetry, e.g. if it was aligned with the ecliptic plane or not.

The above results also clearly indicate that both the detected non-Gaussianity and asymmetry on the large scales cannot mainly be attributed to foreground influences of the Galactic plane. This verifies the results of the full sky surrogate analyses of \cite{Raeth09,Raeth10}, where foreground effects could in principle have played a role, and confirms previous findings of other analyses (e.g. \cite{Hansen09}). In combination with the multitude of checks on systematics performed for the cut sky transformation in section \ref{validationsection} and for the technique of surrogates in \cite{Raeth09,Raeth10}, the results point towards the conclusion that the signatures are of cosmological origin. This would represent a strong violation of the Gaussian hypothesis and of statistical isotropy. Both assumptions are fundamental parts of single-field slow-roll inflation, which is therefore rejected at high significance by this analysis.

A lot of effort as well as knowledge about the cut sky transformation process will be necessary to be able to extend the method of cut sky surrogates to higher resolutions. A naive approach results in an $O(\ell^4)$ scaling. As usual in CMB analysis, this includes both calculations in Fourier as well as in real space. It is important to remove the restrictions of a low $\ell_{max}$ to enable investigations of higher $\ell$-ranges, as it was already done for the full sky case \cite{Raeth09,Raeth10}. When going to higher $\ell$-values, it is of course helpful to work with a higher resolution in pixel space as well. The upcoming data sets of the Planck satellite will offer a better resolution for both Fourier and pixel space.

\section{Conclusions}

We demonstrated the feasibility of generating surrogates by Fourier-based methods also for an incomplete data set. This was worked out for the case of a CMB analysis on an incomplete sphere. Three different constant latitude sky cuts were applied. For this purpose, three different cut sky transformations were calculated. We generated 100 cut sky surrogates for every input map, sky cut and matrix decomposition method, which were analysed by means of scaling indices and Minkowski functionals. To remove systematic effects, a second analysis compared the results of the original with the ones of 20 full sky surrogate maps for each of the input maps. For simulated maps, no anomalies could be detected. The findings for the data maps show strong signatures of non-Gaussianities and pronounced asymmetries, which persist even when removing larger parts of the sky. This confirms that the influence of the Galactic plane is not responsible for these deviations from Gaussianity and isotropy. Together with former full-sky analyses, the results point towards a violation of statistical isotropy. Similar tests with the forthcoming Planck data will yield more information about the origin of the detected anomalies.

Many of the results in this paper have been obtained using HEALPix \cite{Gorski05}. We acknowledge the use of LAMBDA. Support for LAMBDA is provided by the NASA Office of Space Science.


\label{lastpage}

\end{document}